\documentclass[a4paper,twocolumn,superscriptaddress,8pt,
accepted=2019-12-28]{quantumarticle}
\pdfoutput=1
\usepackage[numbers,sort&compress]{natbib}
\usepackage{graphicx}
\usepackage{amsmath}
\usepackage{amssymb}

\usepackage{color}

\usepackage[english]{babel}
\usepackage{subfig}
\usepackage[T1]{fontenc}
\usepackage{mathrsfs}
\usepackage[colorlinks]{hyperref}

\begin{document}

\title{Control of anomalous diffusion of a Bose polaron}

\author{Christos Charalambous}
\affiliation{ICFO -- Institut de Ci\'{e}ncies Fot\'{o}niques, The Barcelona Institute
of Science and Technology, 08860 Castelldefels (Barcelona), Spain}

\author{Miguel \'{A}ngel Garc\'{\i}a-March}
\affiliation{ICFO -- Institut de Ci\'{e}ncies Fot\'{o}niques, The Barcelona Institute
of Science and Technology, 08860 Castelldefels (Barcelona), Spain}
\affiliation{Instituto Universitario de Matem\'atica Pura y Aplicada, Universitat Polit\`ecnica de Val\`encia,  E-46022 Val\`encia, Spain}

\author{Gorka Mu\~noz-Gil}
\affiliation{ICFO -- Institut de Ci\'{e}ncies Fot\'{o}niques, The Barcelona Institute
of Science and Technology, 08860 Castelldefels (Barcelona), Spain}

\author{Przemys{\l}aw Ryszard Grzybowski}
\affiliation{Faculty of Physics, Adam Mickiewicz University, Umultowska 85, 61-614 Pozna{\'n}, Poland}

\author{Maciej Lewenstein}
\affiliation{ICFO -- Institut de Ci\'{e}ncies Fot\'{o}niques, The Barcelona Institute
of Science and Technology, 08860 Castelldefels (Barcelona), Spain}
\affiliation{ICREA, Lluis Companys 23, E-08010 Barcelona, Spain}

\begin{abstract}
We study the diffusive behavior of a Bose polaron immersed
in a coherently coupled two-component Bose-Einstein Condensate (BEC). We assume a uniform,  one-dimensional BEC. %We find that the impurity can have a transient subdiffusive behavior in this system. 
Polaron superdiffuses if it couples in the same manner to both components, i.e.
either attractively or repulsively to both of them. This is the same behavior as that of an impurity immersed in a single BEC. Conversely, the polaron exhibits a transient nontrivial subdiffusive behavior if it couples attractively to one of the components and repulsively to the
other. The anomalous diffusion exponent and the duration of the subdiffusive interval can be controlled
with the Rabi frequency of the coherent coupling between the two
components, and with the coupling strength of the impurity to the BEC.  
\end{abstract}

 \maketitle

\section{Introduction}

The phenomenon of anomalous diffusion attracts a growing interest in classical and quantum physics, appearing in a plethora of various systems~\citep{2005Hanggi,2005Sokolov}. 
In classical systems, there has been a considerable effort to elucidate
the properties and conditions of anomalous diffusive behavior, with
a large emphasis given to the question of how this anomalous diffusion could potentially
be controlled. In many models, the appearance of the anomalous
diffusion is  attributed to some random component of the system-environment
setup, usually distributed with a power-law. Examples include continuous
time random walks~\citep{1975Scher}, diffusion on a fractal lattice~\citep{1994Bunde}, diffusivity (i.e. diffusion coefficient)
that is inhomogeneous in time~\citep{1993Saxton,1997Saxton}, or space~\citep{1986Leyvraz,2012Hottovy,2013aCherstvy,2013bCherstvy,2014Cherstvy} in a regular or random manner,
the patch model~\citep{2014Massignan,2015Manzo}, hunters model~\citep{2017Charalambous}, etc. 
%
%With the radical increase of experimental techniques that allow us to manipulate matter down to the meso and micro scales, interest is nowadays focusing on a microscopic mechanism to explain the appearance of such anomalous diffusion in quantum systems. 
%
In quantum systems, a paradigmatic instance of a highly controlled 
system is that of a Bose Einstein Condensate (BEC). It was shown that BEC with tunable
interactions, are promising systems to study a number of diffusion-related
phenomena, such as Anderson Localization (AL) in disordered media~\citep{2012Min,2008Roati,2012Jendrzejewski}, the expansion of 1D BEC in disordered speckle potentials~\citep{2010SanchezPalencia,2010Modugno,2008Billy,2008Roati,2012Jendrzejewski,2010Deissler,2011Lucioni,2012Min,2017Donsa},
the subdiffusive behavior of the expansion of a wave packet of a 1D quantum, chaotic and nonlinear
system~\citep{1993Shepelyansky,2008Kopidakis,2008Pikovsky,2009Flach,2009Skokos,2009Veksler,2010Mulansky,2010Laptyeva,2010Iomin,2009Larcher,2011Lucioni,2012Min}, the Brownian motion of solitons in BEC~\citep{2017Aycock}, as well as the superdiffusive motion of an impurity in a BEC studied in \citep{2017Lampo,2018Lampo,2019Charalambous}. 

%Last but not least, in .. the authors studied the non-equilibrium dynamics of an impurity suddenly immersed in a BEC until the final formation of a Bose polaron. 

In this work, we study how  an impurity
in a coherently
coupled two-component BEC shows a transient anomalous diffusing behavior. We study this phenomenon under  experimentally relevant conditions, as long as the BEC can be approximated as uniform and one dimensional. We show that this transient anomalous diffusing behavior can be controlled through the strength of the interactions and the coherent coupling. To this end, we treat the Bose Polaron problem within an open quantum system framework. The open quantum system approach  has been used recently in the context of ultracold quantum gases to study the diffusion of an impurity and two impurities in a BEC \citep{2017Lampo,2018Lampo,2019Charalambous}, for the movement of a bright soliton in a superfluid in one dimension \citep{2016Efimkin}, see also\citep{2017Hurst,2018Keser,2012Bonart}). On the other hand, the effect of contact interactions, dipole-dipole interactions
and disorder on the diffusion properties of 1D dipolar two-component
condensates were studied in~\citep{2015XiaoDong}, identifying again
the conditions for subdiffusion. The study of the diffusive
behavior of a 2D two-component BEC in a disordered potential was undertaken in~\citep{2014KuiTian}. Finally, an important study on   an impurity immersed in a two-component BEC was reported in \citep{2018Ashida}. 

The most important novel result of this work, is that we show that under certain assumptions, one can observe a transient subdiffusive behavior of the immersed impurity. We study this for experimentally feasible parameters, as long as the BEC can be approximated as uniform and one dimensional, and we examine how the strength of the coherent coupling and interactions modify this subdiffusive behavior.

To be more specific about the particularities of the system considered in this work, we assume that  an external field drives the population transfer (spin-flipping) between the two atomic levels. The population transfer between the two levels turns out to be described by Josephson dynamics, leading to what is known as internal Josephson effect (see e.g.~\citep{2001Leggett}).
 This internal Josephson interaction controls the many-body physics
of multicomponent phase coherent matter. Importantly, after diagonalizing the Hamiltonian through a Bogoliubov transformation, one obtains a spectrum that has two branches: the density mode, ungapped and  with a linear behavior at low momenta; and the spin mode, gapped, with a parabolic behavior even at low momenta.

From the technical point of view, we identify how under suitable assumptions, starting from the Hamiltonian
describing the aforementioned system of an impurity in a coherently
coupled two component BEC, one can equivalently describe the impurity
as a Brownian particle in a bath, where the role of the bath is played
by the Bogoliubov modes of the coherently coupled two-component BEC. Furthermore, we show that the two branches obtained after the Bogoliubov transformation, the density mode and the spin mode, mentioned above, result in two distinct spectral densities, which we derive in Section~\ref{sec:SD}. We consider two scenarios: same coupling among the impurity and the two bosonic components, and repulsive coupling to one component and attractive to the other. We show that these scenarios 
correspond to the impurity coupling either to the density or to the
spin mode of the two-component BEC, respectively. For the coupling to the density mode there is no qualitative difference in comparison to the
case where the impurity is embedded in a single BEC~\citep{2017Lampo}. For the coupling to the spin mode, we find a different spectral density, namely a gapped sub-ohmic spectral density. We derive and solve the equations of motion of the impurity. These are obtained through the corresponding Heisenberg equations for the bath and impurity
particles and they have the form of Generalized Langevin equations with
memory effects. By solving numerically these equations we find 
the effect of the gapped sub-Ohmic spectral density on the Mean Square
Displacement (MSD) of the impurity.  

The paper is organized as follows. In Section~\ref{sec:Hamiltonian} we introduce the model Hamiltonian and transform it into the form of a Caldeira-Leggett  one. In Section ~\ref{sec:SD} we derive the spectral densities for the cases of coupling to the density or spin modes. In Section~\ref{sec:Heqs} we find and solve the Langevin equations and in Section~\ref{sec:results} we present the results. We end the paper with the discussion and outlook presented in Section~\ref{sec:conclusions}. 

\section{Hamiltonian}
\label{sec:Hamiltonian}

\begin{figure}[h]
%\centering{}%
%\begin{tabular}{|c|}
%\hline 
\includegraphics[width=1\columnwidth]{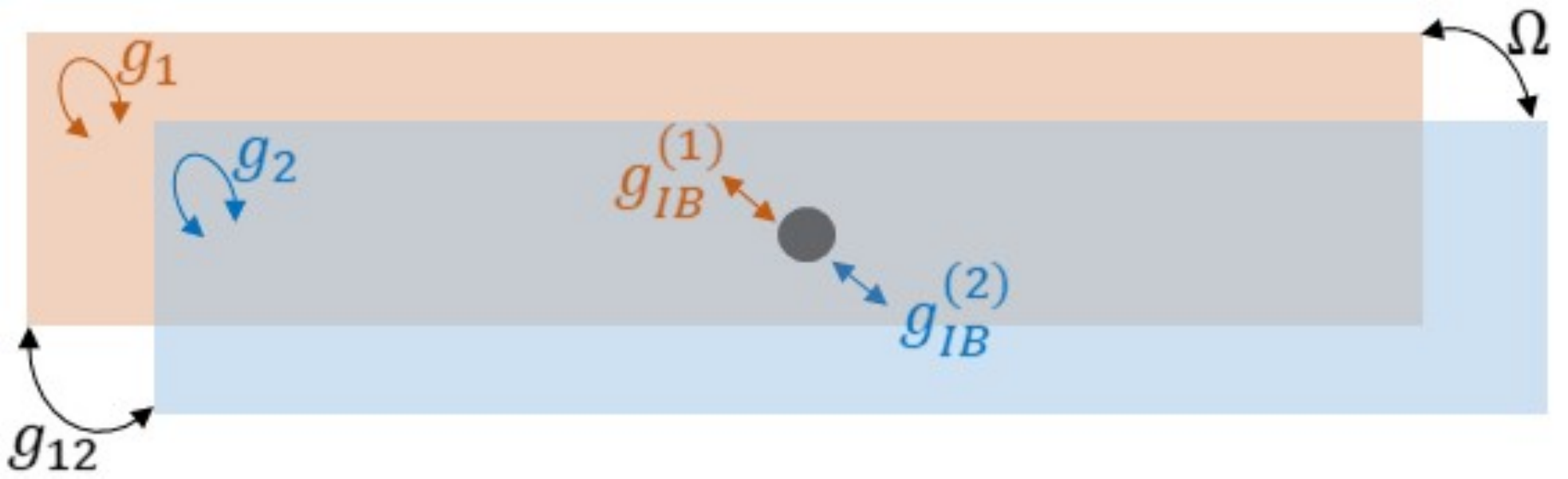}
%\tabularnewline
%\hline 
%\end{tabular}
\caption{\label{fig:fig0} We consider a setup of a coherently coupled two-component BEC in which an impurity is immersed. By $g_1$ and $g_2$ we denote the intraspecies contact interactions of the atoms of the first and second species respectively. $g_{12}$ refers to the coupling strength of the interspecies contact interaction among atoms of the first and second species. By $\Omega$ we denote the Rabi frequency of the Raman coherent coupling between the two species. Finally $g_{IB}^{(1)}$ and $g_{IB}^{(2)}$ indicate the coupling of the impurity to the atoms of the first and second species respectively. }
\end{figure}

The dilute Bose-Einstein condensates created in atomic gases \citep{1995Davis}
consist of bosons with internal degrees of freedom: the atoms can be trapped in different atomic hyperfine states. Soon after
the first observation of atom trap BECs, experimentalists succeeded
in trapping partly overlapping BECs of atoms in different hyperfine
states that are (i) hyperfine split~\citep{1997Myatt} or (ii) nearly
degenerate and correspond to different orientations of the spin \citep{1998StamperKurn}.
We consider a two-component Bose gas with both one-body
(field-field) and two-body (density-density) couplings, composed of
such atoms in different hyperfine states. Furthermore, we assume that
the two components are coupled through a Josephson (one-body) type of
coupling. The two-body interaction results from short-range particle-particle
interactions between atoms in different internal states, while the
one-body interaction can be implemented by two-photon Raman optical
coupling, which transfers atoms from one internal state to the other.
In present-day BEC experiments, the internal Josephson or Rabi interactions,
interconverting atoms of different internal states, consist in two-photon
transitions, induced by a laser field or a combination of a laser field
and oscillating magnetic field. To gain a perspective on the experimental
relevance of our study, we refer the readers to the work of Refs.~\citep{1999Miesner,1998Stenger,1998Matthews,1997Myatt}.
Finally, we assume an impurity that is immersed in the two-component BEC. This impurity interacts with both components through contact interactions. In Fig.~\ref{fig:fig0} a sketch of the setup is shown.
 
 The Hamiltonian of an impurity interacting with a two-species bosonic
mixture in one dimension reads 
\begin{align}
H=H_{\mathrm{I}}+H_{\mathrm{B}}^{(1)}+H_{\mathrm{B}}^{(2)}+H_{\mathrm{IB}}+H_{\mathrm{B}}^{(12)},
\end{align}
where the impurity of mass $m_{\mathrm{I}}$ is described by $H_{\mathrm{I}}=\frac{ \textbf{p}^{2}}{2m_{\mathrm{I}}}+U(\textbf{x})$,
with $U(\textbf{x})$ being the trapping potential. The interactions with
the bosons are described by $H_{\mathrm{IB}}$. We study here only the case of free impurities,
hence we assume $U(\textbf{x})=0$. The terms of the individual bosonic species, labeled with the index $j=1,2$, are  
\begin{align*}
H_{\mathrm{B}}^{(j)}&=\int\Psi_{j}^{\dagger}\left(\textbf{x}\right)\left[-\frac{\textbf{p}_{j}^{2}}{2m^{j}_{\rm{B}}}+V^{j}\left(\textbf{x}\right)\right]\Psi_{j}\left(\textbf{x}\right)\mathrm{d}\textbf{x}\\
&+\frac{g_{j}}{2}\int\Psi_{j}^{\dagger}\left(\textbf{x}\right)\Psi_{j}^{\dagger}\left(\textbf{x}\right)\Psi_{j}\left(\textbf{x}\right)\Psi_{j}\left(\textbf{x}\right)\mathrm{d}\textbf{x},
\end{align*}
where the intra-species contact interactions have a strength given by
the coupling constant $g_{j}=4\pi \hbar^2 a_{\mathrm{B}}^{(j)}/m_{\rm{B}}^{j}$, with $a_{\mathrm{B}}^{(j)}$ the scattering length for the $j^{th}$ species atoms, $m_{\rm{B}}^{j}$ the mass of the atoms of these species and the external
potential for the atoms of the $j^{th}$ species is denoted by $V^{j}\left(x\right)$. For simplicity, we assume $V^{1}\left(x\right)=V^{2}\left(x\right)=V\left(x\right)$ and $m^{1}_{\rm{B}}=m^{2}_{\rm{B}}=m_{\rm{B}}$. Furthermore we focus on the idealized case of an untrapped bath i.e. $V\left(x\right)=0$ which results in a homogeneous density for the BEC. This is not a physical scenario, but in practice, for a BEC trapped in a large box, the impurity in the middle of this box would indeed approximately interact with a bath of constant density. 

The coupling Hamiltonian between the two bosonic species consists in inter-species contact interactions, with coupling constant
$g_{12}$, and a Rabi coupling $\Omega$, which exchanges atoms between
components, i.e., 
\begin{align}
H_{\mathrm{B}}^{(12)} & =g_{12}\int\Psi_{1}^{\dagger}\left(\textbf{x}\right)\Psi_{2}^{\dagger}\left(\textbf{x}\right)\Psi_{2}\left(\textbf{x}\right)\Psi_{1}\left(\textbf{x}\right)\mathrm{d}\textbf{x}\nonumber \\
 & + \hbar {\bf \Omega} \int\Psi_{1}^{\dagger}\left(\textbf{x}\right)\Psi_{2}\left(\textbf{x}\right)\mathrm{d}\textbf{x}+\mathrm{H.c}.
\end{align}
where $g_{12}=4\pi \hbar^2 a_{\mathrm{B}}^{(12)}/m_{\rm{B}}$, with $a_{\mathrm{B}}^{(12)}$ the scattering length for the intraspecies interactions. 
Without any loss of generality, we will only consider $\bf{\Omega}$ real and
positive. This is because even if a complex Rabi frequency is assumed, this can always be cancelled by introducing a counteracting phase for one of the BECs, which can be shown to have no effect on the energy spectrum of the bath.  The latter part of the Hamiltonian, referred to as an internal Josephson interaction, is a two-photon
transition that is induced by a laser field or a combination of a
laser field and an oscillating (rf) magnetic field. This also introduces
an effective energy difference between the two internal states/species
of the BEC, which, assuming a low intensity driving field, is simply
equal to the detuning $\delta$ of the two-photon transition. This
detuning does not affect our studies however, so for sake of clarity
and simplicity, we will assume it to be zero. We also consider here only  repulsive two-body coupling, i.e. $g_{12}>0$.  

In what concerns
the impurity-bosons interaction part of the Hamiltonian, we assume
that the interaction is between the impurity and the densities of the
bosons, i.e. it has the form of a contact interaction:
\begin{equation}
H_{\mathrm{IB}}=\left(\sum_{j=1,2}g_{\rm{IB}}^{(j)}\Psi_{j}^{\dagger}(\textbf{x})\Psi_{j}(\textbf{x})+h.c.\right).
\end{equation}
with $g_{\mathrm{IB}}^{(j)}=2\pi \hbar^2 a_{\mathrm{IB}}^{(j)}/m_{\rm{R}}$, where
$a_{\mathrm{IB}}^{(j)}$
represents the scattering length of the impurity with the bosons of
the $j^{th}$ BEC, and $m_{\rm{R}}=m_{\rm{B}}m_{\rm{I}}/(m_{\rm{B}}+m_{\rm{I}})$ is the relevant
reduced mass.
From this point onwards, we assume that the BEC is one dimensional, which simplifies the analytical part of our studies. Nevertheless, the main result of our work, which is the control of the dynamics of an impurity in a coherently coupled two-component BEC by coupling it either to the density mode of the BEC or to the spin mode of the BEC, will remain irrespective of the dimension. The diffusing behavior itself however might change, since the spectral density depends on the dimension of the BEC. It is worth noting here, that the 1D case in fact is peculiar since, in principle, in the purely 1D scenario (not in the confined 3D elongated/cigar shaped case) the condensation is destroyed by the phase fluctuations \cite{2000Petrov}. However, if the phase coherence length is larger than the band-size, then one can speak about "true" BEC. In other words, as long as the physics of interest happens on the scales smaller than the phase coherence length, it is legitimate to use BEC and Bogolubov-de Gennes theory, as we will do in our work. 

Furthermore, we comment here that, had we considered the more realistic case of a harmonically trapped BEC, assuming it to be described by a Thomas-Fermi density profile, this would lead to a discretized Bogoliubov spectrum, which would then have to be treated under a similar approximation as in the single BEC case in \citep{2018Lampo}. In principle, we expect that the maim result of this work, i.e. being able to couple the impurity to two distinct types of baths, should hold, but the related diffusing behaviours might change. We expect that when the gap of the spin models is much bigger than the trap frequency, the sub-diffusion, that we observe in Sec. \ref{sec:results}, should persist in the same form. The reader should also have in mind that, as discussed in the appendix of \citep{2018Lampo} the limit of the trapping frequency of the BEC going to 0 would not lead to the homogeneous BEC case. This scenario requires a more careful treatment, that goes beyond the scope of our paper.

A further assumption we make in our studies is that of a dilute gas of low depletion, since in this case, we will be able to apply the Bogoliubov diagonalization technique and obtain the energy spectrum of the BEC bath. In the low-density sub-milikelvin temperature regime of the atom
trap experiments, we may assume that the trapped atoms interact only through
the partial $s$-wave channel, and that the many-body properties are well
described by assuming the particles to interact as hard spheres. The
radius of those spheres is given by the scattering length $a$, which we assume to be positive. We
say that the system of particle density $n$ is dilute if the packing
fraction  of space occupied by the spheres $na^{3}\ll1$. The assumption of low depletion means that almost all particles
occupy, on average, the single particle state associated with the
condensate ($k=0$, where $k$ is the momentum for the particular case
of homogeneous BEC that we will be considering). This implies that the temperatures to be considered should be smaller
than the critical temperature.

For a single BEC, all the
bosons condensate at the same state. However, this will not be the
case for the two-component BEC and one has to determine the fraction of particles in each component, which will depend on the ground state of the system. This is determined by the parameters of the system. 

With the above considerations in mind, we assume that the two bosonic gases
condense. This means that we can apply mean field theory
and further assuming that the ground state is coherent, the wavefunctions
$\Psi_{j}\left(x\right),\Psi_{j}^{\dagger}\left(x\right)$ for a homogeneous
BEC are given by 
\begin{equation}
\Psi_{j}(x)=\Psi_{j,0}(x)+\delta\Psi_{j}(x),
\end{equation}
where $\Psi_{j,0}(x)=\phi_{0}(x)\sqrt{N_{j}}e^{i\theta_{j}}$, with
$\theta_{j}$ being the phase of the coherent $j^{th}$ component, $N_{j}$ the number of bosons of the $j^{th}$ species  
and $\delta\Psi_{j}(x)=\sum_{k\neq0}\phi_{j,k}(x)a_{j,k}$ with $\phi_{k}(x)=\frac{1}{\sqrt{V_{j}}}e^{ikx/\hbar}$
the plane wave solutions, with $V_{j}$ the corresponding bath's volume.
From here onwards we assume for simplicity that $V_{1}=V_{2}=V$, i.e.
that the two baths have the same volume, and that we are dealing with
homogeneous BECs.  Here $a_{j,k} $ and $ a_{j,k}^{\dagger}$ are bosonic annihilation and creation operatos.  To proceed further, we write the
Hamiltonian in terms of these operators. The bosonic
parts read 
\begin{align*}
H_{\mathrm{B}}^{(j)}&=\sum_{k}\epsilon_{k}a_{j,k}^{\dagger}a_{j,k}\nonumber\\
&+\frac{g_{j}}{2}\sum_{k,k',q}a_{j,k+q}^{\dagger}a_{j,k'-q}^{\dagger}a_{j,k}a_{j,k'},\\
H_{\mathrm{B}}^{(12)}&=g_{12}\!\!\!\!\sum_{k,k',q}\!\!\!a_{1,k+q}^{\dagger}a_{2,k'-q}^{\dagger}a_{2,k}a_{1,k'}\nonumber\\
&+\hbar \Omega\!\sum_{k}a_{1,k}^{\dagger}a_{2,k}\!+\!\mathrm{H.c.},
\end{align*}
with $\epsilon_{k}=k^{2}/2m_{\rm{B}}$.  The zeroth order expectation value (or mean field value) of the Hamiltonian reads as
\begin{align}
H_0 &= 
\sum_{j} \frac{g_j}{2V} \Psi_{j,0}^4+\frac{g_{12}}{V}\Psi_{1,0}^2\Psi_{2,0}^2\nonumber \\ 
&+\hbar \Omega\left(\Psi_{1,0}(\Psi_{2,0})^*+(\Psi_{1,0})^*\Psi_{2,0}\right)\label{eq:zeroth order Ham}
\end{align}

%From now onwards, we set $\hbar=1$ and we adimensionalized correspondingly. Note that we also  assume $m_{I}=1$ from now onwards.

\subsection{Generalized Bogoliubov transformation}
\label{sec:Gen Bog Trans}

We now  perform a generalized
Bogoliubov transformation to diagonalize the Bosonic part of the Hamiltonian and hence obtain the energy spectrum of the bath. We follow closely the results of \cite{2003Tommasini,2013Lellouch,2013Abad} in the rest of this section. This generalized
Bogoliubov transformation is understood to be composed of a rotation,
a scaling and one more rotation as in \citep{2003Tommasini}. The
derivation is based on a simple geometrical picture which results
in a convenient parametrization of the transformation. %In addition, the simplicity of the geometrical picture on which the construction of the Bogoliubov transformation is based upon indicates that this procedure may be extended to higher component BECs.
% We note here that
%one can also  use the phase-density Bogoliubov-Popov
%approach, which allows to treat true condensates and quasi-condensates \citep{1972Popov,1983Popov}, which would be more
%rigorous since we are dealing with a quasi-condensate (a 1D BEC). 
%Anyhow, 
%the results we present below would be unaffected by this. 
%Finally,
%the tools of the emerging BEC technology have been used to directly
%probe elementary excitations, for single BEC as in \citep{1999StamperKurn,1997Timmermans,2002Vogels},
%and have hence allowed for a test for the quasiparticle description
%resulting from a Bogoliubov transformation. The same measurement,
%can similarly probe the many-body structure of a two-component BEC
%and provide a detailed test of the related Bogoliubov theory that
%is used in what follows. 
Following the generalized Bogoliubov transformation, the initial bath operators are transformed as 
\begin{align}
a_{j,k} & =\mathcal{Q}^0_{j,+,k}\,b_{+,k}+\mathcal{Q}^1_{j,+,k}\,b_{+,-k}^{\dagger}\nonumber\\
& +\left(-1\right)^{\delta_{j,-}}\left(\mathcal{Q}^0_{j,-,k}\,b_{-,k}+\mathcal{Q}^1_{j,-,k}\,b_{-,-k}^{\dagger}\right)\nonumber\\
a_{j,-k}^{\dagger}& =\mathcal{Q}^1_{j,+,k}\,b_{+,k}+\mathcal{Q}^0_{j,+,k}\,b_{+,-k}^{\dagger}\label{eq:spinsensityoperators}\\
&+\left(-1\right)^{\delta_{j,-}}\left(\mathcal{Q}^1_{j,-,k}\,b_{-,k}+\mathcal{Q}^0_{j,-,k}\,b_{-,-k}^{\dagger}\right),\nonumber
\end{align}
with $b_{+(-),k}^\dagger$ and $b_{+(-),k}$ and  the creation/annhilation operators for the final spin ($+$) and density or phonon  ($-$) mode. In the latter,  the total density fluctuates, while in the spin mode $(+)$ the unlike particle densities fluctuate out of phase.
This, in the presence of an internal Josephson interaction as
in our case, is a Josephson plasmon \cite{2001Paraoanu}. In Eq.~\eqref{eq:spinsensityoperators} the parameters are as follows:  $ \delta_{1(2),-(+)}=1$, 
$\delta_{1(2),+(-)}=0$ and
\begin{align}
&\mathcal{Q}^\phi_{j,s,k}=\label{eq:Q} \\
&R_{sj}\hat\Gamma_{j,s,k}\Big[\left(1\!-\!\delta_{j,s}\right)\cos\left(\gamma_{k}\right)\!+\!\delta_{j,s}\sin\left(\gamma_{k}\right)\Big]\cos\theta\nonumber\\
&\!+\!R_{sj'}\hat\Gamma_{j',s,k}\Big[\!\left(1\!-\!\delta_{j',s}\!\right)\cos\left(\gamma_{k}\right)\!+\!\delta_{j',s}\sin\left(\gamma_{k}\!\right)\!\Big]\!\sin\theta, \nonumber
\end{align}
where $j'\neq j$, $\phi \in \left\lbrace 0,1 \right\rbrace$, $s \in \left\lbrace +,- \right\rbrace$, $\hat\Gamma_{j(j'),s,k}=[\Gamma_{j(j'),s,k}^{2}+\left(-1\right)^{\phi}]/2\Gamma_{j(j'),s,k} $ and $R_{1+}=R_{2-}=\left(1, -1\right)^\top$, $R_{2+}=-R_{1-}=\left(1,
1\right)^\top$. Also,  in Eq.~(\ref{eq:Q}), 
\begin{align}
&\sin\left(\gamma_{k}\right)=\\
&\sqrt{\frac{1}{2}\left[1-\frac{[\omega_{1,k}^{2}-\omega_{2,k}^{2}]}{\sqrt{(\omega_{1,k}^{2}-\omega_{2,k}^{2})^{2}+16\Lambda_{12}^{2}n_{1}n_{2}e_{1,k}e_{2,k}}}\right]},\nonumber
\end{align}
with $\cos\left(\gamma_{k}\right)$ defined accordingly, and $\Gamma_{j,s,k}=\sqrt{e_{j,k}/E_{s,k}}$
where 
\begin{align}
&E_{\pm,k}:=\hbar\Omega_{\pm,k}=\label{Eq:Spectrum}\\
&\left[\!\frac{\sum_{j}\!\omega_{j,k}^{2}\!\pm\!\sqrt{(\omega_{1,k}^{2}-\omega_{2,k}^{2})^{2}\!+\!16\Lambda_{12}^{2}n_{1}n_{2}e_{1,k}e_{2,k}}}{2}\right]^{\!\frac{1}{2}}\!\!,\nonumber
\end{align}
with
\begin{align}
e_{j,k}&=\epsilon_{k}\!-(-1)^j\!\frac{\left(1\!+(-1)^j\!\cos{\theta_{12}}\right)\!\hbar\Omega n}{2n_{1}n_{2}}\nonumber\\
\omega_{j,k}&=\sqrt{e_{j,k}^{2}+2\Lambda_{j}n_{j}e_{j,k}},\nonumber\\
\Lambda_{1}n_{1}&=g_{1}n_{1}\cos^{2}\left(\theta\right)\nonumber\\
& + g_{2}n_{2}\sin^{2}\left(\theta\right)\!+\!g_{12}\sin\left(2\theta\right)\cos\left(\theta_{12}\right)\!,\nonumber\\
\Lambda_{2}n_{2}&=g_{1}n_{1}\sin^{2}\left(\theta\right)\nonumber\\
&+g_{2}n_{2}\cos^{2}\left(\theta\right)\!-\!g_{12}\sin\left(2\theta\right)\cos\left(\theta_{12}\right)\!,\nonumber\\
\Lambda_{12}\!\sqrt{n_{1}n_{2}}&=\frac{g_{2}n_{2}\!-\!g_{1}n_{1}}{2}\!\sin\!\left(\!2\theta\right)\nonumber\\
&+g_{12}\sqrt{n_{1}n_{2}}\!\cos\left(2\theta\right)\!\cos\!\left(\!\theta_{12}\!\right)\!,\nonumber
\end{align}
where $n_{j}=\frac{N_{j}}{V}$ is the particle density of the $j^{th}$
bath, and $\theta$ is the free parameter (angle) to be determined by the minimisation of the total energy.  In the above expression, $\theta_{12}=\theta_{1}-\theta_{2}$
is the relative phase between the two BEC. The minimization of the zeroth energy with respect to the angle $\theta$ gives 
\begin{equation}
\tan(\theta)=\begin{cases}
\sqrt{\frac{n_{1}}{n_{2}}}, & \text{if }\theta_{12}=\pi,\\
\sqrt{\frac{n_{2}}{n_{1}}}, & \text{if }\theta_{12}=0.
\end{cases}
\end{equation}

\begin{figure}[h!]
%\centering{}%
%\begin{tabular}{|c|}
%\hline 
\includegraphics[width=0.9\columnwidth]{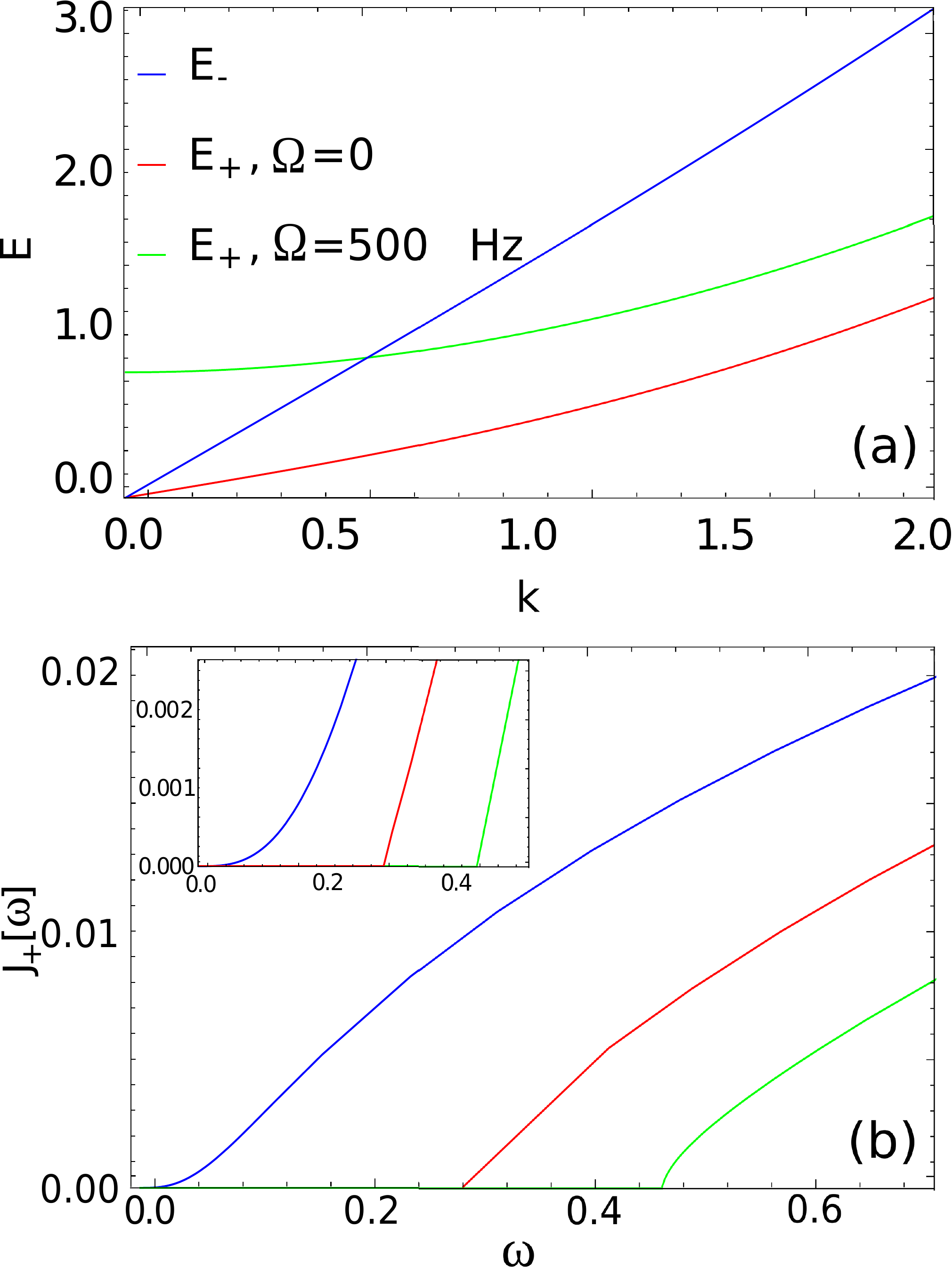}
%\tabularnewline
%\hline 
%\end{tabular}
\caption{\label{fig:fig1} (a) Energy spectrum for a coherentely coupled two component BEC.  There are  two branches in the spectrum corresponding to the  density (-) and spin modes (+). We plot both for different values of the coherent coupling, $\Omega$. First, this illustrates that the gap opens for the spin mode (+); and second, it shows that, while for $\Omega=0$ both branches behave similarly, i.e., linearly for low $k$ and quadratically for large $k$, for finite $\Omega$ the (+) mode behaves quadratically even at low $k$. This has direct implications on the behavior of the spectral density in case 2, plotted in (b). When the $\Omega=0$ (blue line) the
spectral density behaves as for the density mode (i.e. with a $~w^3$-behavior). The red  and green lines (for $\Omega=50\pi, 100\pi$Hz,   respectively) show instead a different behavior. The inset shows a zoom, where we checked that it fits the simplified behavior in Eq.~\eqref{eq:approximate spectral density}, i.e. has a lower gap and behaves as $\sqrt{\omega}$ initially. In these plots we used $g=g_{12}=2.15\times10^{-37}J\cdot m$ , $n=7(\mu m)^{-1}$ , $g_{\rm{IB}}=0.5\times10^{-37}J\cdot m$, with  BEC and  impurities made of Rb and K atoms, respectively. }
\end{figure}

%\begin{figure}[h]
%%\centering{}%
%%\begin{tabular}{|c|}
%%\hline 
%\includegraphics[width=0.9\columnwidth]{Fig1.eps}
%%\tabularnewline
%%\hline 
%%\end{tabular}
%\caption{\label{fig:fig1} (a) Energy spectrum for a coherentely coupled two component BEC.  There are  two branches in the spectrum corresponding to the  density (-) and spin modes (+). We plot both for different values of the coherent coupling, $\Omega$. First, this illustrates that the gap opens for the spin mode (+); and second, it shows that, while for $\Omega=0$ both branches behave similarly, i.e., linearly for low $k$ and quadratically for large $k$, for finite $\Omega$ the (+) mode behaves quadratically even at low $k$. This has direct implications on the behavior of the spectral density in case 2, plotted in (b). When the $\Omega=0$ (blue line) the
%spectral density behaves as for the density mode (i.e. with a $~w^3$-behavior). The red  and green lines (for $\Omega=50\pi, 100\pi$Hz,   respectively) show instead a different behavior. The inset shows a zoom, where we checked that it fits the simplified behavior in Eq.~\eqref{eq:approximate spectral density}, i.e. has a lower gap and behaves as $\sqrt{w}$ initially. In these plots we used $g=g_{12}=2.15x10^{-37}J\cdot m$ , $n=7\mu m$ , $g_{\rm{IB}}=0.5x10^{-37}J\cdot m$, with  BEC and  impurities made of Rb and K atoms, respectively. }
%\end{figure}

By minimizing
with respect to the population imbalance $f=\frac{N_{1}-N_{2}}{N}$,
one can obtain the following conditions on the parameters of the system
in order to have an extremum of the energy, 
\begin{equation}
\Delta+Af-\cos\theta_{12}\frac{f}{\left(1-f^{2}\right)^{1/2}}=0,\label{eq:equilibrium condition}
\end{equation}
where $A=\frac{\left(g_{1}+g_{2}-2g_{12}\right)n}{4\hbar \Omega}$ is the mutual
interaction parameter, $\Delta=\frac{2\delta+\left(g_{1}-g_{2}\right)n}{4\hbar\Omega}$ is
the effective detuning parameter, and
\begin{equation}
A-\cos\theta_{12}\frac{1}{\left(1-f^{2}\right)^{3/2}}>0,
\end{equation}
is the condition to have a minimum of the energy. In \citep{2003Tommasini},
it was shown that to obtain the minimum energy of the
system, without imposing any condition on the detuning $\delta$,
as is our case, then the relative phase should be chosen to be $\theta_{12}=\pi$,
referred to as the $\pi$-state configuration. %Furthermore, we will need in our work, to be able to invert the spectrum $E_{\pm,k}$, in order to obtain an expression for the momentum $k$ in terms of the frequencies of the bath \MAGM{what?}. This will allow to integrate over this momenti\MAGM{what?} in an expression for a quantity called spectral density, and hence obtain an analytic expression for it, which will then allow us to characterize the diffusive behavior of the impurity. To do this inversion, we will need to resort to a specific case for our system
From here on we assume the symmetric case, i.e,  $g_{1}=g_{2}=g$ as this  will allow us to obtain analytically the spectral density in section \ref{sec:SD}. In this case, the equilibrium condition Eq. \eqref{eq:equilibrium condition}
reads as
\begin{equation}
\left(g-g_{12}+\frac{\hbar\Omega}{\sqrt{n_{1}n_{2}}}\right)\left(n_{1}-n_{2}\right)=0,
\end{equation}
which has two solutions 
\begin{equation}
\begin{array}{cc}
n_{1}-n_{2}=0 & \left(\mbox{GS1}\right),\\
n_{1}-n_{2}=\pm n\sqrt{1-\left(\frac{2\hbar\Omega}{\left(g-g_{12}\right)n}\right)^{2}} & \left(\mbox{GS2}\right),
\end{array}
\end{equation}
corresponding to neutral GS1  and polarized ground
states GS2. Here, we make the strong Josephson junction assumption 
\begin{equation}
\left|A\right|<1,\label{eq:MISCIBILITY CONDITION}
\end{equation}
which is also referred to as the {\it miscibility condition}. 
This implies that the minimum energy equilibrium ground state has to be GS1  as is shown in \citep{2003Tommasini,2013Abad}. Handable expressions for the spectral density obtained  section \ref{sec:SD} are possible over GS1. For the regime in which ground state is GS2 we expect similar qualitative behavior, but we did not obtained a form for the spectral density which allows us to obtain the diffusive behavior of the impurity. The study of the  impurity diffusion over GS2, and even at the phase transition, falls out of the scope of this paper. 
Under the miscibility condition~\eqref{eq:MISCIBILITY CONDITION}, the energy spectrum expressions simplifies into
\begin{align}
E_{-,k}\! =&\left(\epsilon_{k}\left(\epsilon_{k}+\left(g+g_{12}\right)n\right)\right)^{\frac{1}{2}}, \label{eq:simplified spectrum-}\\
E_{+,k}\!=&\left[\epsilon_{k}\left(\epsilon_{k}+\left(g-g_{12}\right)n+4\hbar \Omega\right)\right.\nonumber\\
&\left.  + 2\hbar \Omega\left[\left(g-g_{12}\right)n+2\hbar \Omega\right]\right]^{\frac{1}{2}}.\label{eq:simplified spectrum}
\end{align}

In Fig.~\ref{fig:fig1} we plot the energy  spectra as a function of $k$ for specific parameters, to illustrate the spin and density branches.  Furthermore,
note that the bogoliubov transformation elements satisfy the well-known
relation
\begin{equation}
\mathcal{Q}^0_k(\mathcal{Q}^0_k)^{T}-\mathcal{Q}^1_k(\mathcal{Q}^1_k)^{T}=1,
\end{equation}
where $\mathcal{Q}^{\phi}_{k}=\left(\begin{array}{cc}
\mathcal{Q}^{\phi}_{1,+,k} & \mathcal{Q}^{\phi}_{1,-,k}\\
\mathcal{Q}^{\phi}_{2,+,k} & \mathcal{Q}^{\phi}_{2,-,k}
\end{array}\right)$ with $\phi \in \left\lbrace 0,1 \right\rbrace$, that
implies normalization. However, as is shown in \citep{2013Lellouch},
these Bogoliubov operators, do not fulfill the bosonic commutations
relations, which is understood as a consequence of the fact that they
are not orthogonalized with respect to the quasicondensate functions
$\Psi_{j,0}=\sqrt{N_{j}}e^{i\theta_{j}}$. In \citep{2013Lellouch} it is  shown that
to overcome this problem one needs to define some new transformation
with components $\widehat{\mathcal{Q}}^0_{j,s,k},\widehat{\mathcal{Q}}^1_{j,s,k}$, that
are related to the previous ones as
\begin{equation}
\widehat{\mathcal{Q}}^{\phi}_{j,s,k}=\mathcal{Q}^{\phi}_{j,s,k}-\frac{\Psi_{j,0}}{N_{j}}\mathcal{Q}^{\phi}_{j,s,k}\Psi_{j,0}^{\ast}.
\end{equation}
The elements of this transformation, these new Bogoliubov operators,
are expressed in terms of the Bogoliubov wave functions of our system
$f_{j,s,k}$, $\widetilde{f}_{j,s,k}$ as 
\begin{equation}
\widehat{\mathcal{Q}}^{\phi}_{j,s,k}=\frac{f_{j,s,k}+\left(-1\right)^{\phi}\widetilde{f}_{j,s,k}}{2},
\end{equation}
where 
\begin{align*}
&f_{1,-,k}=f_{2,-,k}=\left[\frac{\epsilon_{k}}{2E_{-,k}}\right]^{1/2},\\
&\widetilde{f}_{1,-,k}=\widetilde{f}_{2,-,k}=\left[\frac{E_{-,k}}{2\epsilon_{k}}\right]^{1/2},\\
&f_{1,+,k}=f_{2,+,k}=\left[\frac{\epsilon_{k}+\hbar \Omega}{2E_{+,k}}\right]^{1/2},\\
&\widetilde{f}_{1,+,k}=\widetilde{f}_{2,+,k}=\left[\frac{E_{+,k}}{2\left(\epsilon_{k}+\hbar \Omega\right)}\right]^{1/2}.
\end{align*}

The spin mode branch is gapped while the density mode branch is gapless.
For the latter, at low values of the momentum $k$ the dispersion
is linear, with a speed of sound $c_{d}=\sqrt{n\left(g+g_{12}\right)/(2m_{\rm{B}})}$.
On the contrary for the gapped branch, the dispersion relation goes
as $k^{2}$ for low $k$, and at $k=0$, it has a gap 
\begin{align}
E_{\mathrm{gap}}=\sqrt{2\hbar \Omega\left[\left(g-g_{12}\right)n+2\hbar \Omega\right]}.\label{eq:energy gap}
\end{align}
This corresponds to the Josephson frequency for small amplitude oscillations.
As we will see the fact that there are two branches in the spectrum
will give rise to two different noise sources. 

Furthermore, one should note that,  had we not introduced
the Rabi coupling term in the Hamiltonian, the latter would commute
with both $n_{1}$ and $n_{2}$ such that we would have two broken
continuous symmetries and both branches would be gapless (notice that $E_{\mathrm{gap}}\rightarrow0$ when
$\hbar \Omega\rightarrow0$). In this case, the low momentum excitations
would be both phase-like, as it has to be for Goldstone modes of the
$U\left(1\right)$x$U\left(1\right)$ broken symmetries \citep{2019Recati}. Hence the introduction of the Rabi coupling
term, results in the system having only one continuous broken symmetry,
namely only $n$ has to be conserved now and not both $n_{1}$ and
$n_{2}$. The long wavelength limit of the Goldstone mode corresponds
to a low-amplitude phonon fluctuation in which the total density oscillates
and the unlike atoms move in unison (i.e., with the same superfluid
velocity). In contrast, in the long wavelength gap mode the unlike
atoms move in opposite directions, while their center of mass remains
at rest. This fluctuation is then reminiscent of the motion of ions
in an optical phonon mode, which also exhibits a gapped dispersion.
At zero momentum, the gap mode corresponds to an infinitesimal Josephson-like
oscillation of the populations in the distinguishable internal states.
In the strong Josephson coupling regime we have closed orbits around
a fixed point for the Josephson Hamiltonian, with vanishing mean polarisation
(or population imbalance) and a phase difference around $\pi$ if
$\hbar \Omega\geq0$, giving rise to plasma-like oscillations.

\subsection{Transformed Impurity-Bath interaction}
\label{sec:Imp-Bath int}

In terms of the original annihilation and creation operators,  the impurity-bath term reads as 
\begin{align}
H_{\mathrm{IB}} & =\sum_{j}\frac{1}{V}\sum_{k,q}V_{\mathrm{IB}}^{(j)}\left(k\right)\rho_{\mathrm{I}}\left(q\right)a_{j,k-q}^{\dagger}a_{j,k}\label{Eq:Interacting Hamiltonian}\\
 & =\sum_{j}\sqrt{\frac{n_{j}}{V}}\sum_{k\neq0}\rho_{\mathrm{I}}\left(k\right)V_{\mathrm{IB}}^{(j)}\left(a_{j,k}+a_{j,-k}^{\dagger}\right),\nonumber 
\end{align}
with $\rho_{I}\left(q\right)=\int_{-\infty}^{\infty}e^{-iqx'}\delta\left(x'-x\right)dx'$,
$V_{\mathrm{IB}}^{(j)}\left(k\right)=\mathcal{F}_{k}[g_{\mathrm{IB}}^{(j)}\delta\left(x-x'\right)]$ where
$\mathcal{F}$ is the Fourier trasnform. Furthermore, $n_{j}$, with $j=1,2$, is the averaged density
of the $j^{th}$ bath. The second line of Eq.~\eqref{Eq:Interacting Hamiltonian}
is a consequence of the assumption that the bosons condense. We
will consider two cases for the coupling of the impurity to the baths.
\begin{enumerate}
\item{In the first scenario the impurity couples to the two baths in the same way
\begin{equation}
\, g_{\rm{IB}}^{\left(1\right)}=g_{\rm{IB}}^{\left(2\right)}=g_{\rm{IB}},
\end{equation}}
\item{while in the second scenario, the interactions are attractive with one of the baths and repulsive with the other the other, 
\begin{equation}
\,  g_{\rm{IB}}^{\left(1\right)}=-g_{\rm{IB}}^{\left(2\right)}=g_{\rm{IB}}. 
\end{equation}}
\end{enumerate}

After the Bogoliubov transformation the impurity-bath term reads in both  cases as  
\begin{align*}
1.\, H_{IB}^{\left(-\right)}\!=\!\left[\frac{n}{V}\right]^{\!\!\frac{1}{2}}\!\!\!\sum_{j,k\neq0}\!\!\rho_{I}\left(k\right)g_{\rm{IB}}\left[\widehat{\mathcal{Q}}^0_{j,-,k}\!+\!\widehat{\mathcal{Q}}^1_{j,-,k}\right]\!x_{-,k}, 
\end{align*}
\begin{align*} 
2.\,H_{IB}^{\left(+\right)}\!=\!\left[\frac{n}{V}\right]^{\!\!\frac{1}{2}}\!\!\!\sum_{j,k\neq0}\!\!\rho_{I}\left(k\right)g_{\rm{IB}}\left[\widehat{\mathcal{Q}}^0_{j,+,k}\!+\!\widehat{\mathcal{Q}}^1_{j,+,k}\right]\!x_{+,k},
\end{align*}
where $x_{\pm,k}:=\left(b_{\pm,k}+b_{\pm,k}^{\dagger}\right)$. These equations show that in case 1   the impurity only couples to the density $(-)$ mode of the bosonic baths, while in case 2 it couples only to the spin $(+)$ mode. 
For both cases, we rewrite the impurity-bath terms as 
\begin{align}
H_{\mathrm{IB}}^{s}=\sum_{\substack{j,k\neq0\\
s\in\left\{ +,-\right\}
}
}V_{j,s,k}e^{ikx}\left(b_{s,k}+b_{s,-k}^{\dagger}\right),\label{Eq:Himp-bath}
\end{align}
where $s=-$ for case 1 and $s=+$ for case 2, and
\begin{equation}
V_{j,s,k}=\sqrt{\frac{n}{V}}g_{\mathrm{IB}}\left(\widehat{\mathcal{Q}}^0_{j,s,k}\!+\!\widehat{\mathcal{Q}}^1_{j,s,k}\right).\label{Eq:crucialEq}
\end{equation}
We note here that $\widehat{\mathcal{Q}}^0_{1,s,k}\!+\!\widehat{\mathcal{Q}}^1_{1,s,k}=\widehat{\mathcal{Q}}^0_{2,s,k}\!+\!\widehat{\mathcal{Q}}^1_{2,s,k}$, such that $V_{1,s,k}=V_{2,s,k}=\widehat{V}_{s,k}$. We linearize the interaction (see~\citep{2017Lampo} for validity
of this assumption) to get 
\begin{equation}
H_{\mathrm{IB}}=\sum_{\substack{k\neq0\\s\in\left\{ +,-\right\} }} V_{s,k}\left(\mathbb{I}+ikx\right)\left(b_{s,k}+b_{s,-k}^{\dagger}\right).\label{eq:Linearized Ham}
\end{equation}
where $V_{s,k}=2\widehat{V}_{s,k}$. Thus, after a redefinition $b_{s,k} \rightarrow b_{s,k}-\frac{V_{s,k}}{E_{s,k}} \mathbb{I}$, the final total Hamiltonian  reads as 
\begin{equation}
H=H_{\mathrm{I}}+\sum_{\substack{k\neq0\\
s\in\left\{ +,-\right\} 
}
}E_{s,k}b_{s,k}^{\dagger}b_{s,k}+ \sum_{\substack{k\neq0\\
s\in\left\{ +,-\right\} 
}
} \hbar g_{s,k}\pi_{s,k},\label{Eq:caldeira-Leggett}
\end{equation}
with $g_{j,s,k}=kV_{s,k}/\hbar$ and $\pi_{s,k}=i\left(b_{k,s}-b_{k,s}^{\dagger}\right)$
the momentum of the bath particles. We see that as in ~\citep{2017Lampo},
the coupling occurs between the position of the impurity and the momentum
of the bath particles. However, in our work, the coupling can take place to one of the two different quasiparticle branches, depending on the form of the impurity-baths interactions.

\section{Spectral densities}
\label{sec:SD}
The spectral densities can be obtained from the self-correlation functions~\citep{2017Lampo} for each environment (corresponding to cases 1 and 2). These  read
\begin{equation}
\mathcal{C}\left(t\right)=\!\!\!\!\!\!\! \sum_{\begin{array}{c}
k\neq0\\
s\in\left\{ +,-\right\} 
\end{array}}\!\!\!\!\!\!\!\hbar g_{s,k}^{2}\left\langle \pi_{s,k}\left(t\right)\pi_{s,k}\left(0\right)\right\rangle/\hbar.
\end{equation}
Using that the bath is composed of bosons for which
\begin{equation}
\left\langle b_{k,s}^{\dagger}b_{k,s}\right\rangle =\frac{1}{e^{\frac{\hbar \omega_{k}}{k_{B}T}}-1},
\end{equation}
we obtain 
\begin{align}
&\mathcal{C}\left(t\right)\!=\!\!\!\!\!\!\!
\sum_{\substack{k\neq0 \\ s\in\left\{ +,-\right\}}}\!\!\!\!\!\!
g_{s,k}^{2}\left[\!\coth\left(\!\frac{\hbar \omega_{k}}{2k_{B}T}\!\right)\!\!\cos\left(\omega_{k}t\right)\!-\!i\sin\left(\omega_{k}t\right)\right]\nonumber\\
&=\nu\left(t\right)-i\lambda\left(t\right),
\end{align}
where 
\begin{align}
\nu\left(t\right)&=\!\!\!\int_{0}^{\infty}\!\!\!\!\!\sum_{\substack{
s\in\left\{+,-\right\} }}\!\!\!\!\!J^{D}\left(\omega\right)\coth\left(\frac{\hbar \omega}{2k_{B}T}\right)\cos\left(\omega t\right)d\omega,\nonumber\\
\lambda\left(t\right)&=\int_{0}^{\infty}\!\!\!\sum_{\substack{
s\in\left\{+,-\right\} }}\!\!\!\!J^{D}\left(\omega\right)\sin\left(\omega t\right)d\omega.
\label{eq:noise and damping kernel}
\end{align}
In these definitions we used the spectral density,  
\begin{equation}
J^{D}\left(\omega\right)= \hbar \sum_{k\neq0}\left(g_{s,k}\right)^{2}\delta\left(\omega-\omega_{k}\right).
\end{equation}
The spectral density is then evaluated in the continuous frequency limit as
\begin{align}
&J^{D}\left(\omega\right)=\\
&4ng_{\rm{IB}}^{2}\frac{D_{d}}{\hbar\left(2\pi\right)^{d}}\!\!\!\int\!\! dk k^{d\!+\!1}\!\left(\mathcal{U}_{s,k}\!\!+\!\!\mathcal{V}_{s,k}\right)^{2}\!\!\frac{\delta\left(k\!-\!k_{E_{s}}\!\left(\!\omega\!\right)\right)}{\partial_{k}E_{s}\!\left(k\right)\!\mid_{k=k_{E_{s}}\!\left(\!\omega\!\right)}},\nonumber
\end{align}
where $D_{d}$ is the surface of the hypersphere in the momentum space
with radius $k$ in $d$-dimensions. In the particular case of
1D becomes $D_{1}=2$. 

To obtain the expression for the
continuous frequency case, the inverse of the dispersion
relation from Eq.~\eqref{Eq:Spectrum} is needed. For this general energy spectrum, obtaining such inverse function is not easy. However this is indeed possible for the simplified
case, Eqs.~\eqref{eq:simplified spectrum-}--\eqref{eq:simplified spectrum}.   The inverse of the density
(-) branch which is the one to which the impurity couples for the
case 1 type of coupling, reads 
\begin{align}
&k_{E_{-}}\left(\omega\right)=\\
&\sqrt{m_{\rm{B}}ng}\left[\!\frac{g_{12}}{g}\!-\!1+\sqrt{1\!-\!\frac{2g_{12}}{g}\!+\!\left[\frac{g_{12}}{g}\right]^{2}\!+\!\left[\frac{2\omega}{ng}\right]^{2}}\right]^{\!\!\frac{1}{2}}\!\!\!.\nonumber
\end{align}
%It is furthermore, important to note that the spectrum of the two modes, i.e. the density and spin modes, are different in character at low values of the momenta $k$. In particular, the density spectrum is linear and gapless as was the case of the spectrum for the single BEC, but the spin spectrum goes as $k^{2}$ and in addition is gapped.
With this, for the density ($-$) branch (case 1 type of coupling), the spectral density is 
\begin{equation}
J_{-}\left(\omega\right)=\tilde{\tau}_{-}\frac{G_{-}(\omega)^{3/2}}{\sqrt{F_{-}(\omega)}},
\end{equation}
with 
\begin{align}
F_{-}(\omega) & =1+\left(\frac{\omega}{\Lambda_{-}}\right)^{2},\\
G_{-}(\omega) & =-1+\sqrt{F_{-}(\omega)},\\
\tilde{\tau}_{-} & =\frac{\left(2g_{\rm{IB}}\right)^{2}nm_{\rm{B}}^{3/2}}{2^{1/2}\pi}\sqrt{\Lambda_{-}},
\end{align}
and where $\Lambda_{-}=n\left(g+g_{12}\right)/2\hbar$ is the cutoff frequency,
which resembles the one in \citep{2017Lampo} when $g$ is replaced
by $\frac{g+g_{12}}{2}$. In the limit of $\omega\ll\Lambda_{-}$, the spectral
density can be simplified to 
\begin{equation}
J_{-}\left(\omega\right)=m_I \tau_{-}\omega^{3},\label{eq:Density mode spectral density}
\end{equation}
where 
\begin{equation}
\tau_{-}=\frac{\left(2\eta\right)^{2}}{2\pi m_I}\left(\frac{m_{\rm{B}}}{n\left(g+g_{12}\right)^{1/3}}\right)^{3/2},
\end{equation}
with $\eta_{-}=\frac{g_{\rm{IB}}}{g+g_{12}}$. Thus, for the the $\pi$-state
equilibrium configuration one obtains a cubic spectral density. 

\subsection{Spin branch coupling}
\label{sec:Spin branch}

For the spin ($+$) branch (case 2 type of coupling), the inverse
of the spectrum reads as
\begin{align}
&k_{E_{+}}\left(\omega\right)=\sqrt{m_{\rm{B}}ng}\times\\
&\left[\frac{g_{12}}{g}\!-\!1\!-\!\frac{4\hbar \Omega}{ng}+\sqrt{1\!-\!\frac{2g_{12}}{g}\!+\!\left[\frac{g_{12}}{g}\right]^{2}\!+\!\left[\frac{2\omega}{ng}\right]^{2}}\right]^{\frac{1}{2}}.\nonumber
\end{align}
In this case, the spectral density is
\begin{equation}
J_{+}\left(\omega\right)=m_I\widetilde{\tau}_{+}\frac{G_{+}\left(\omega\right)}{\sqrt{F_{+}\left(\omega\right)}},
\label{eq:Spin spectral density-expr}
\end{equation}
where 
\begin{align}
F_{+}\left(\omega\right)&=1+\left(\frac{\omega}{\Lambda_{+}}\right)^{2},\nonumber\\
G_{+}\left(\omega\right)&\!=\!W\!\left(\omega\right)\!\left[W\!\left(\omega\right)\!+\!\frac{1}{2}\!-\!\frac{1}{2}\sqrt{1\!+\!\left[\frac{E_{\mathrm{gap}}}{\Lambda_{+}}\right]^{2}}\right]^{\!\frac{1}{2}}\!,\nonumber\\
W\left(\omega\right)&=-\frac{1}{2}+\sqrt{F_{+}\left(\omega\right)}-\frac{1}{2}\sqrt{1+\left[\frac{E_{\mathrm{gap}}}{\Lambda_{+}}\right]^{2}},\nonumber\\
\widetilde{\tau}_{+}&=\frac{\left(2g_{\rm{IB}}\right)^{2}nm_{\rm{B}}^{3/2}}{2^{1/2}\pi m_I}\sqrt{\Lambda_{+}},
\label{eq:Spin spectral density}
\end{align}
with $\Lambda_{+}=n\left(g-g_{12}\right)/2\hbar$. We note that  to interpret $\widetilde{\tau}_{+}$
as a relaxation time, as is custom to do (see \citep{2017Lampo}),  one has to impose  $g\geq g_{12}$ to assure it remains  a real quantity. In other case, the spectral density will be imaginary (note $\frac{G_{+}\left(\omega\right)}{\sqrt{F_{+}\left(\omega\right)}}$
is independent of the sign of $g-g_{12}$).

Let us find how the spectral density in Eq. \eqref{eq:Spin spectral density-expr} simplifies in two limiting cases. First, in the absence of coherent coupling,  $\Omega=0$, the gap vanishes, $E_{\mathrm{gap}}=0$. In this case, Eq. \eqref{eq:Spin spectral density-expr}  is equal to that of the density mode, upon the  interchange  $\Lambda_{-}\rightarrow\Lambda_{+}$. Therefore, on the  long time limit $\omega\ll\Lambda_{+}$ we obtain the same cubic behavior of the spectral density. We illustrate this case in Fig. \ref{fig:fig1}. In panel (a) we show  that the two branches of the energy spectra  have the same behavior, that is, linear at low $k$ and parabolic for large $k$. 

Second,  we consider the case
of finite $\Omega$ which implies $E_{\mathrm{gap}}>0$. A  requirement which we  impose on the spectral
density is that one cannot consider frequencies lower than
the gap energy $E_{\mathrm{gap}}$. Physically, one can interpret this as follows: Since the energy spectrum of the bath is gapped, with a gap given by Eq. \eqref{eq:energy gap},  the spectral density
cannot assign a weight at frequencies lower than this, because the
bath cannot excite the impurity with such frequencies since it is
not part of its spectrum. Then, we simplify the spectral density as 
\begin{equation}
\widehat{J}_{+}\left(\omega\right)=\Theta\left(\omega-E_{\mathrm{gap}}\right)J_{+}\left(\omega\right),
\end{equation}
with $\Theta(.)$ the step delta function. 

 Let us now  comment on the frequency region right above the energy
gap of our system.  To this end, we replace $\omega=E_{\mathrm{gap}}+\epsilon$, where
$\epsilon>0$. We use $\epsilon$ as the small value expansion parameter
in our case, i.e. we consider the limit $\epsilon\ll E_{\mathrm{gap}}$, such that $\omega\approx E_{\mathrm{gap}}$. We furthermore introduce
a cutoff $\Lambda$, for which it holds that $\epsilon\ll\Lambda-E_{\mathrm{gap}}$.
The expressions in the spectral density will now read as
\begin{align}
&F_{+}\left(\epsilon\right)=1+\left(\frac{E_{\mathrm{gap}}}{\Lambda_{+}}\right)^{2},\\
&W\left(\epsilon\right)=\nonumber \\
&-\!\frac{1}{2}\!+\!\frac{1}{2}\sqrt{\!1\!+\!\left(\frac{E_{\mathrm{gap}}}{\Lambda_{+}}\right)^{2}}\!+\!\frac{E_{\mathrm{gap}}}{\Lambda_{+}}\left( E_{\mathrm{gap}}^{2}+\Lambda_{+}^{2} \right)^{-\frac{1}{2}}\epsilon,\nonumber
\end{align}
such that 
\begin{align}
&G_{+}\!\left(\epsilon\right)=\\
&\left[\!-\frac{1}{2}\!+\!\frac{1}{2}\!\sqrt{\!1\!+\!\!\left[\frac{E_{\mathrm{gap}}}{\Lambda_{+}}\right]^{2}}\right]\!\!\left[\!\frac{E_{\mathrm{gap}}}{\Lambda_{+}}\!\right]^{\!\!\frac{1}{2}}\!\left[E_{\mathrm{gap}}^{2}\!+\!\Lambda_{+}^{2} \right]^{\!-\!\frac{1}{4}}\!\epsilon^{\frac{1}{2}}.\nonumber
\end{align}
Hence the spectral density is
\begin{equation}
\widehat{J}_{+}\left(\epsilon\right)=\Theta\left(\epsilon\right)\tau_{+}\epsilon^{1/2},
\end{equation}
with 
\begin{equation}
\tau_{+}=\widetilde{\tau}_{+}\frac{\left(-\frac{1}{2}+\frac{1}{2}\sqrt{1+\left(\frac{E_{\mathrm{gap}}}{\Lambda_{+}}\right)^{2}}\right)\left(E_{\mathrm{gap}}\right)^{\frac{1}{2}}}{\left(1+\left(\frac{E_{\mathrm{gap}}}{\Lambda_{+}}\right)^{2}\right)^{1/4}}.
\end{equation}
The final form of the spectral density, after introducing a cutoff
$\Lambda$, to avoid the related ultraviolet divergencies mentioned above, is
\begin{align}
&\widehat{J}_{+}\!\left(\omega\right)=\label{eq:approximate spectral density}\\
&\Theta\left(\omega-E_{\mathrm{gap}}\right)\tau_{+}\left[\omega-E_{\mathrm{gap}}\right]^{\frac{1}{2}}\Theta\left(\Lambda+E_{\mathrm{gap}}-\omega\right).\nonumber
\end{align}
We introduced a hard cutoff to our spectral density as this
better describes the physical system we study. Such a spectral density,
i.e. with an exponent on the frequencies less than 1, is often associated  to subdiffusive impurity dynamics. Note that in the results that
we present below, we assume $g\approx g_{12}$ as in \citep{2013Nicklas},
which significantly simplifies the expression for the coefficient
of the spectral density $\tau_{+}$ without changing its
behavior. In particular in this case $\tau_{+}=
\left(2g_{\rm{IB}}\right)^{2}nm_{\rm{B}}^{3/2}/2^{1/2}\pi m_I$.
In Fig. \ref{fig:fig1}(b), we show how  the approximate spectral density in Eq.
\eqref{eq:approximate spectral density}, under the assumption $\epsilon\ll E_{\mathrm{gap}}$,
compares to the original spectral density. For  vanishingly small
values of $E_{\mathrm{gap}}$ the spectral density approaches the
form of that of the density mode, i.e. it goes as $\propto\omega^{3}$, as expected. In the limit we are interested, that is, for finite $E_{\mathrm{gap}}$, the spectral density  behaves approximately as in Eq.~\eqref{eq:approximate spectral density} (see inset in Fig.~\ref{fig:fig1}(b)). 

Such gapped spectral densities as in Eq.~\eqref{eq:approximate spectral density},
have been studied extensively in the literature. In general,
they are usually related to semiconductors \citep{1994John,2011Tan,2013Prior}
or photonic crystals (PC) \citep{1994Kofman}. In particular, the simplified form of the spectral
density, Eq.~\eqref{eq:approximate spectral density}, is related in particular with 3D PCs. The latter,
are artificial materials engineered with periodic dielectric structures
\citep{1972Bykov}. If one considers an atom embedded in such a material,
it is known that if the resonant frequency of the excited atom approaches
the band gap edge of the PC, strong localization of light, atom-photon
bound states, inhibition of spontaneous emission and fractionalized
steady-state inversion appear \citep{1987Yablonovitch,2000Lambropoulos,2003Woldeyohannes,1997Quang}.
The rapidly varying distribution of field modes near the band gap
\citep{1994Kofman,2004Kofman} requires a non-Markovian
description \citep{2007Breuer} of the reduced dynamics of quantum
systems coupled to the radiation field of a PC \citep{2003Woldeyohannes,1997Quang,2010Rivas,2005deVega,2008deVega}.
This enhanced appearance of non-Markovian effects is also confirmed
by a recent study based on exact diagonalization \citep{2014Vasile}.
In this study it was observed that for frequencies of the bath much
larger than the band gap, energy transfer between the system and the
bath is such that information and energy flow irreversibly
from system to bath leading to Markovian dynamics. Conversely, at the edges of the gaps, one observes the largest backflow
of information where the energy bounces between the system and bath
leading to non-Markovian evolution of the system. Furthermore, deep
within the band gap, less excitations and energy are exchanged between
the system and the bath, which is shown to lead to localized modes
\citep{2012Zhang}, expressed as dissipationless oscillatory behavior,
plus non-exponential decays (such as for example fractional relaxations
\citep{2014Giraldi}). 

From the work in \citep{2012Zhang}, a relationship is suggested of
such long-lived oscillations that appear in the dynamics of a system
coupled to a bath with a gapped spectral density, with the fact that
the Hamiltonian of the system might have thermodynamic and dynamic
instabilities. This is the case when the Hamiltonian is unbounded
from below, i.e. non-positive. Physically this happens when one deals with Hamiltonians that do not conserve the particles number, and this is indeed the case for the Hamiltonian of the free quantum Brownian motion we study here. Hence, as a result, this unbounded Hamiltonian induces dynamical instabilities in the
long-time regime, corresponding to the limit
$\omega\ll\Lambda$. In practice, as in the spirit of \citep{2017Lampo}, one can show that the effect
of the bath on the impurity is not only to dissipate its energy, but
as well to introduce an inverse parabolic potential in which the impurity
is diffusing (which would work as a renormalization of the trapping
frequency had we considered a harmonic trapping potential). This
inverse parabolic potential is understood to be a consequence of the
unboundedness of the Hamiltonian, and is what is resulting in the
dynamical instability of the long time solution of the impurity dynamics.
Unfortunately, contrary to the case in \citep{2017Lampo}, we will
not consider a harmonic trap for the impurity, and hence the positivity
of the Hamiltonian is violated irrespective of the strength of the
coupling of the impurity to the bath. In practice one would study
the impurity constrained in a box of a certain size, which if included
in the modelling of the system, would result in a positively defined
Hamiltonian, at the price of complicating significantly the analytical
solution for the impurity's dynamics. Hence, we assume here that we are looking at timescales
where the effect of the finite sized box are not manifested. Theoretically,
there are also a number of other ways to circumvent this problem,
even without referring to the presence of a box, such as taking into
account bilinear terms in the impurity's or bath's operators in the
Hamiltonian as in \citep{2007Bruderer,2013Rath,2015Christensen,2016Shchadilova}.
In any case, we will show below that for the regime of the transient
effect that we are interested in, this will not change our results. 

Following the approach sketched above, we take advantage of the simplicity
of the Fr\"ohlich like Hamiltonian we are considering above. Furthermore, we remind that we will look at the long time dynamics of the impurity, i.e. $\omega\ll\Lambda$, as was implied by the above study on the spectral density's form.
In addition, one should take into account the dissipationless oscillatory
behavior, which can also be expressed as an incomplete decay of the
Green function impurity propagator which we  will study below. In
fact by identifying the equivalence of the appearance of these oscillations
with the incomplete decay of the Green function, this provides us
with a very simple condition upon which the long-lived oscillations
appear, that is that the Green function has at least one purely imaginary
pole, which can be shown to only be possible for frequencies within
the band gap \citep{1994Kofman}. We will study this in the next section. 

%Finally, an important point to note, is that from the above considerations,
%we conclude also that our system could potentially be used to observe
%various phenomena traditionally linked to particles immersed in electromagnetic
%fields in quantum-optical systems. This idea goes in parallel to a
%previous attempt to simulate quantum optical phenomena with cold atoms
%in optical lattices as in \citep{2011NavarreteBenlloch}. It is yet
%important to state that this setup could as well be used in simulating
%Bose-Hubbard Hamiltonian with extended hopping and Ising models with
%long-range interactions.

\section{Heisenberg equations and their solution}
\label{sec:Heqs}
In this section, we derive the equation of motion for the impurity, which will allow us to study its diffusive behavior under  various scenarios. To do so, we begin with the Heisenberg equations of motion for both the impurity and the bath particles. The latter set of equations can be solved, and we use this solution to obtain a Langevin like equation of motion for the impurity. 
The Heisenberg equations for the bath particles are
\begin{align}
\frac{db_{s,k}\left(t\right)}{dt}&=i\left[H,b_{s,k}\left(t\right)\right]\nonumber\\
&=-i\Omega_{s,k}b_{s,k}\left(t\right)-\hbar\sum_{j=1}^{2}g_{s,k}^{\left(j\right)}x\left(t\right),\nonumber\\
\frac{db_{s,k}^{\dagger}\left(t\right)}{dt}&=\frac{i}{\hbar}\left[H,b_{s,k}^{\dagger}\left(t\right)\right]\nonumber\\
&=\frac{i}{\hbar}\Omega_{s,k}b_{s,k}^{\dagger}\left(t\right)-\hbar\sum_{j=1}^{2}g_{s,k}^{\left(j\right)}x\left(t\right),
\end{align}
and for the central particle 
\begin{align}
\hspace{-1cm}\frac{dx\left(t\right)}{dt}&=\frac{i}{\hbar}\left[H,x\left(t\right)\right]=\frac{p\left(t\right)}{m_{I}},\\
\frac{dp\left(t\right)}{dt}\!&=\!\frac{i}{\hbar}\left[H,p\left(t\right)\right]\nonumber\\
&=\frac{i}{\hbar}\left[U\!\left(x\right)\!,\!x\left(t\right)\right]\!-\!\sum_{\substack{k\neq0\nonumber\\
j=\{1,2\}\\
s=\{+,-\}
}
}\!\!\!\!\hbar g_{s,k}^{\left(j\right)}\pi_{s,k}\left(t\right).
\end{align}
Substituting the solutions of the equations of motion for the bath
into that of the central particle, one gets 
\begin{equation}
\ddot{x}\left(t\right)+\frac{\partial}{\partial t}\int\Gamma\left(t-s\right)x\left(s\right)ds=\frac{B\left(t\right)}{m_{I}},\label{eq:equation of motion of impurity}
\end{equation}
where
\begin{align}
&\Gamma\left(\tau\right)=\frac{1}{m_{I}}\int_{0}^{\infty}\frac{\sum_{\substack{j=\{1,2\}\nonumber\\
s=\{+,-\}
}
}J_{s}^{(j)}\left(\omega\right)}{\omega}\cos\left(\omega\tau\right)d\omega,\\
&B\left(t\right)=\\
&\underset{k\neq0}{\sum}i \hbar \!\!\!\underset{\substack{j=\{1,2\}\\
s=\{+,-\}
}
}{\sum}\!\!g_{s,k}^{\left(j\right)}\left(b_{s,k}^{\dagger}\left(t\right)e^{i\omega_{k}t}-b_{s,k}\left(t\right)e^{-i\omega_{k}t}\right)\!,\nonumber
\end{align}
are the damping and noise terms, respectively. Note that in Eq. \eqref{eq:equation of motion of impurity}
we neglected a term $-\Gamma\left(0\right)x\left(t\right)$. This term may introduce dynamic instabilities
in our system in the long time regime. As in~\cite{2017Lampo}, we neglect it as these instabilities are unphysical, that is, will not occur in a physical realization of the system and will only occur in the long time behavior. To be more specific, for the coupling to the density mode this term reads as,
\begin{equation}
\Gamma_{-}\left(0\right)=\tau_-\frac{\Lambda_-^3}{3}.
\end{equation} 
For the coupling to the spin mode this term reads as 
\begin{align}
\Gamma_{+}\left(0\right)&=\tau_{+}\Bigg[-\pi E_{Gap}^{1/2}+2\left(\Lambda+E_{Gap}\right)^{0.5}\\
&\times\left.F_{2,1}\left(-\frac{1}{2},-\frac{1}{2};\frac{1}{2};\frac{E_{Gap}}{\Lambda+E_{Gap}}\right)\right].\nonumber
\end{align}

As in \citep{2017Lampo}, the solution of Eq.~\eqref{eq:equation of motion of impurity}
takes the form 
\begin{align}
&x(t)=\\
&G_{1}(t)x(0)+G_{2}(t)\dot{x}(0)+\frac{1}{m_{I}}\!\int_{0}^{t}\!\!G_{2}(t-s)B(s)ds,\nonumber
\end{align}
with the corresponding Green functions given by 
\begin{align}
\mathcal{L}_{z}\left[G_{1}(t)\right]&=\frac{z}{z^{2}+z\mathcal{L}_{z}\left[\Gamma(t)\right]}=\frac{1}{z+\mathcal{L}_{z}\left[\Gamma(t)\right]},\label{eq:G1 LAPLACE}\\
\mathcal{L}_{z}\left[G_{2}(t)\right]&=\frac{1}{z^{2}+z\mathcal{L}_{z}\left[\Gamma(t)\right]},\label{eq:G2 Laplace}
\end{align} 
where $\mathcal{L}_{z}\left[\cdot\right]$ represents the Laplace transform.
%Then, $G_{1}\left(t\right)$ can be obtained without any reference
%on the specific nature of the bath. Since $G_{1}(t)=1$, it is possible
%to invert the Laplace transform in Eq. \eqref{eq:G1 LAPLACE} directly.
 The expressions for $G_{1}(t)$ and $G_{2}(t)$ depend
on the specific type of bath we consider. For the first scenario (coupling to the density mode), in \citep{2017Lampo} 
 it was found that $\Gamma(t)$  is
\begin{align}
&\Gamma_{-}(t)=\\
&\frac{\tau_{-}}{t^{3}}\left[2\Lambda_{-}t\cos\left(\Lambda t\right)-2\left(2-\Lambda_{-}^{2}t^{2}\right)\sin\left(\Lambda_{-}t\right)\right],\nonumber
\end{align}
and under the assumption of $z\ll\Lambda_{-}$ 
\begin{equation}
\mathcal{L}_{z}\left[\Gamma_{-}(t)\right]=\tau_{-}\Lambda_{-}z+\mathcal{O}\left(z^{2}\right),
\end{equation}
which results in 

\begin{equation}
\mathcal{L}_{z}\left[G_{1}(t)\right]=\frac{1}{\left(1+\Lambda_{-}\tau_{-}\right)z},
\end{equation}
and
\begin{equation}
\mathcal{L}_{z}\left[G_{2}(t)\right]=\frac{1}{\left(1+\Lambda_{-}\tau_{-}\right)z^{2}}.
\end{equation}
Then, one obtains
\begin{equation}
G_{1}(t)=\frac{1}{\left(1+\Lambda_{-}\tau_{-}\right)},
\end{equation}
and
\begin{equation}
G_{2}(t)=\frac{t}{\left(1+\Lambda_{-}\tau_{-}\right)},\label{eq:Green fn density}
\end{equation}
where we see that the Green functions has an identical form to that of \citep{2017Lampo}. More importantly, $G_{2}(t)$ diverges at $t\rightarrow\infty$,
a consequence of the fact that an equilibrium state is not reached
at this limit, as we see in next section. 

For the second case (coupling to the spin mode), as discussed in previous section the spectral density is gapped and given by Eq.~\eqref{eq:approximate spectral density}. To proceed, we first need an expression for the Laplace transform $\mathcal{L}_{z}\left[\Gamma_{+}(t)\right]$
which can be shown to read as 
\begin{align}
&\mathcal{L}_{z}\left[\Gamma_{+}(t)\right]=\!\!\int_{0}^{\infty}\!\!\left[\!\int_{0}^{\infty}d\omega\frac{\widehat{J}_{+}\left(\omega\right)}{\omega}\cos\left(\omega t\right)\right]e^{-zt}dt\nonumber\\
&=z\int_{0}^{\infty}d\omega\frac{\widehat{J}_{+}\left(\omega\right)}{\omega\left(\omega^{2}+z^{2}\right)}\!=\!\frac{\tau_{+}}{z\left(\!E_{gap}^{3}\!+\!E_{gap}z^{2}\!\right)}\!\Lambda^{1.5}
\nonumber\\
&\times\left[\!-\frac{E_{\mathrm{gap}}}{3}\!\left(\!E_{\mathrm{gap}}\!+\!iz\!\right)\!F_{2,1}\!\left(\!1,\frac{3}{2};\frac{5}{2};\!-\!\frac{\Lambda}{E_{\mathrm{gap}}\!-\!iz}\!\right)\!\right.\nonumber\\
&-\frac{E_{\mathrm{gap}}}{3}\left(E_{\mathrm{gap}}-iz\right)F_{2,1}\left(1,\frac{3}{2};\frac{5}{2};-\frac{\Lambda}{E_{\mathrm{gap}}+iz}\right)\nonumber\\
&\left.+\frac{2}{3}\left(E_{\mathrm{gap}}^{2}+z^{2}\right)F_{2,1}\left(1,\frac{3}{2};\frac{5}{2};-\frac{\Lambda}{E_{\mathrm{gap}}}\right)\right],
\label{eq:Laplace damping Spin}
\end{align}
where $F_{2,1}\left(\alpha,\beta;\gamma;z\right)$ is the hypergeometric
function
\begin{equation}
F_{2,1}\left(\alpha,\beta;\gamma;z\right)=\sum_{n=0}^{\infty}\frac{\left(\alpha\right)_{n}\left(\beta\right)_{n}}{\left(\gamma\right)_{n}}\frac{z^{n}}{n!},
\end{equation}
with $\left(\cdot\right)_{n}$ being the Pochhammer symbol. Unfortunately, to
invert the Laplace transform in Eq.~\eqref{eq:G2 Laplace}, given Eq.~\eqref{eq:Laplace damping Spin},
is rather complicated. For this reason we restrain ourselves to study
only the long-time limit, determined by $z\ll\Lambda$. In this case
the inverse Laplace transform of the Green's function in Eq.~\eqref{eq:G2 Laplace},
reads as
\begin{align}
&G_{2}(t)=At,\\
&A\!=\!E_{\mathrm{gap}}^{5}\!\left[\!E_{\mathrm{gap}}^{5}\!+\!\frac{2}{3}E_{\mathrm{gap}}^{2}\Lambda^{1.5}\tau_{+}F_{2,1}\!\left(1,\frac{3}{2};\frac{5}{2};\frac{-\Lambda}{E_{\mathrm{gap}}}\right)\right.\nonumber\\
&-\frac{4}{5}E_{\mathrm{gap}}\Lambda^{2.5}\tau_{+}F_{2,1}\left(2,\frac{5}{2};\frac{7}{2};-\frac{\Lambda}{E_{\mathrm{gap}}}\right)\nonumber\\
&\left.+0.285714\Lambda^{3.5}\tau_{+}F_{2,1}\left(3,\frac{7}{2};\frac{9}{2};-\frac{\Lambda}{E_{\mathrm{gap}}}\right)\right]^{-1},\nonumber
\label{eq: Green fn spin}
\end{align}
which has the same time dependence as in case  1 (coupling to the density mode). We also find $G_1(t)=A$ which is again a constant and hence does not play a role in the time dependence of the MSD which we are interested in. Unfortunately this is as far as we
can get analytically, as contrary to the coupling to the density mode,
even though we have the Green function at hand, using it to obtain
an analytic expression for the MSD of the impurity which is our ultimate
goal is not possible.

Equation~\eqref{eq:Green fn density} as well as Eq.~\eqref{eq:Laplace damping Spin}
have been both obtained at the long time limit, which implied expanding
the Laplace transform of the damping kernel $\mathcal{L}_{z}\left[\Gamma_{-}(t)\right],\mathcal{L}_{z}\left[\Gamma_{+}(t)\right]$
at the first order in $z/\Lambda_{-},z/\Lambda$. In general, one could
have considered higher orders of the aforementioned expansion, but
should then be careful in inverting the Laplace transform to obtain
the Green function in defining the relevant Bromwich integral in the
complex plane in such a way as to not include the roots which correspond
to divergent runaway solutions~\citep{2017Lampo}. Even if one would
do so, the result for the Green function would not change much, and
this can be proven by considering a numerical inversion of the Laplace
transform for the Green function, where the long time limit assumption
is not made. For the coupling to the density mode this was shown using
the Zakian method in~\citep{2017Lampo}. For the spin mode, we checked
this using the same method. Moreover, we contrasted its results to
two other methods for numerically inverting a Laplace transform, in
particular, the Fourier and the Stehfest methods~\citep{2015Wang}.
The Zakian method, gives the inverse of the Laplace transform of a
function $F\left(z\right)$ in the following form
\begin{equation}
f\left(t\right)=\frac{2}{t}\sum_{j=1}^{N}Re\left[k_{j}F\left(\frac{\beta_{j}}{t}\right)\right],
\end{equation}
where $k_{j}$ and $\beta_{j}$ are real and complex constants given
in \citep{2015Wang}. With all of these methods,  the Green function
behaves linearly with time for the range of parameters we considered.
In fact in the numerical results presented in the next section, the
Zakian method was used to obtain the Green function, such that our
results are not restricted just to the long time limit, $z\ll\Lambda$. 

In addition, we are also now in a position to check the presence of
the long-lived oscillations in our system. As was mentioned before,
this can only be the case if the Green function exhibits a purely
imaginary pole, which if it exists, should correspond to a frequency
within the bandgap. As is shown in \citep{2015Lo}, this will be the
case, for the frequency that is a solution of
\begin{equation}
\omega^{2}+\Gamma\left(0\right)-\Delta\left(\omega\right)=0,\label{eq:oscillatory condition}
\end{equation}
where 
\begin{align}
\Delta\left(\omega\right)&:=\mathit{P}\int_{0}^{\infty}\frac{\widehat{J}_{+}\left(\omega'\right)}{\omega-\omega'}d\omega'\nonumber\\
&=-\frac{2\tau_{+}\Lambda^{1.5}F_{2,1}\left(1,\frac{3}{2};\frac{5}{2};-\frac{\Lambda}{E_{\mathrm{gap}}-\omega}\right)}{3\left(E_{\mathrm{gap}}-\omega\right)},
\label{eq:self energy}
\end{align}
is the bath self energy correction, where $\mathit{P}$ denotes the principal
value. One can show that the expression of Eq.~\eqref{eq:self energy}
is always non-positive for $\omega<E_{\mathrm{gap}}$ and hence the
condition in Eq.~\eqref{eq:oscillatory condition} is never satisfied,
such that we do not have to worry about these oscillations in the
transient dynamics that we will study in the next section. 

Finally, one can evaluate the validity of the linearity assumption which allowed us to consider a linear coupling between the BEC and the impurity
(see Hamiltonian in Eq.~\eqref{eq:Linearized Ham}) in terms of the physical parameters of the system. This assumption reads as $kx\ll1$. In~\citep{2017Lampo} it was shown that, as a function of the temperature,  there exist a maximum time for which the linear assumption holds. In the system discussed here, since an 
expression for the MSD cannot be found, for each set of parameters one has to evaluate numerically the long-time behavior of the MDS and determine the maximum time for which the assumption holds. To this end, one has to note that, differently to~\citep{2017Lampo},  for the coupling to the spin mode the momenta grows parabolically with $\omega$ even for small $k$ and there is an energy gap. Then, to evaluate the criteria ($kx\ll1$), one has to use the expression for the energy, Eq.~\eqref{eq:simplified spectrum} together with the numerically evaluated MSD. We checked this condition in the numerical examples presented in next section. A final comment is that, we also made sure that the assumption of a purely 1D BEC was valid, by checking that the spatial extend of the motion did not exceed the phase coherence length  \cite{2000Petrov}.

\begin{figure}[h!]
\centering{}%
%\begin{tabular}{|c|}
%\hline 
\includegraphics[width=1.2\columnwidth]{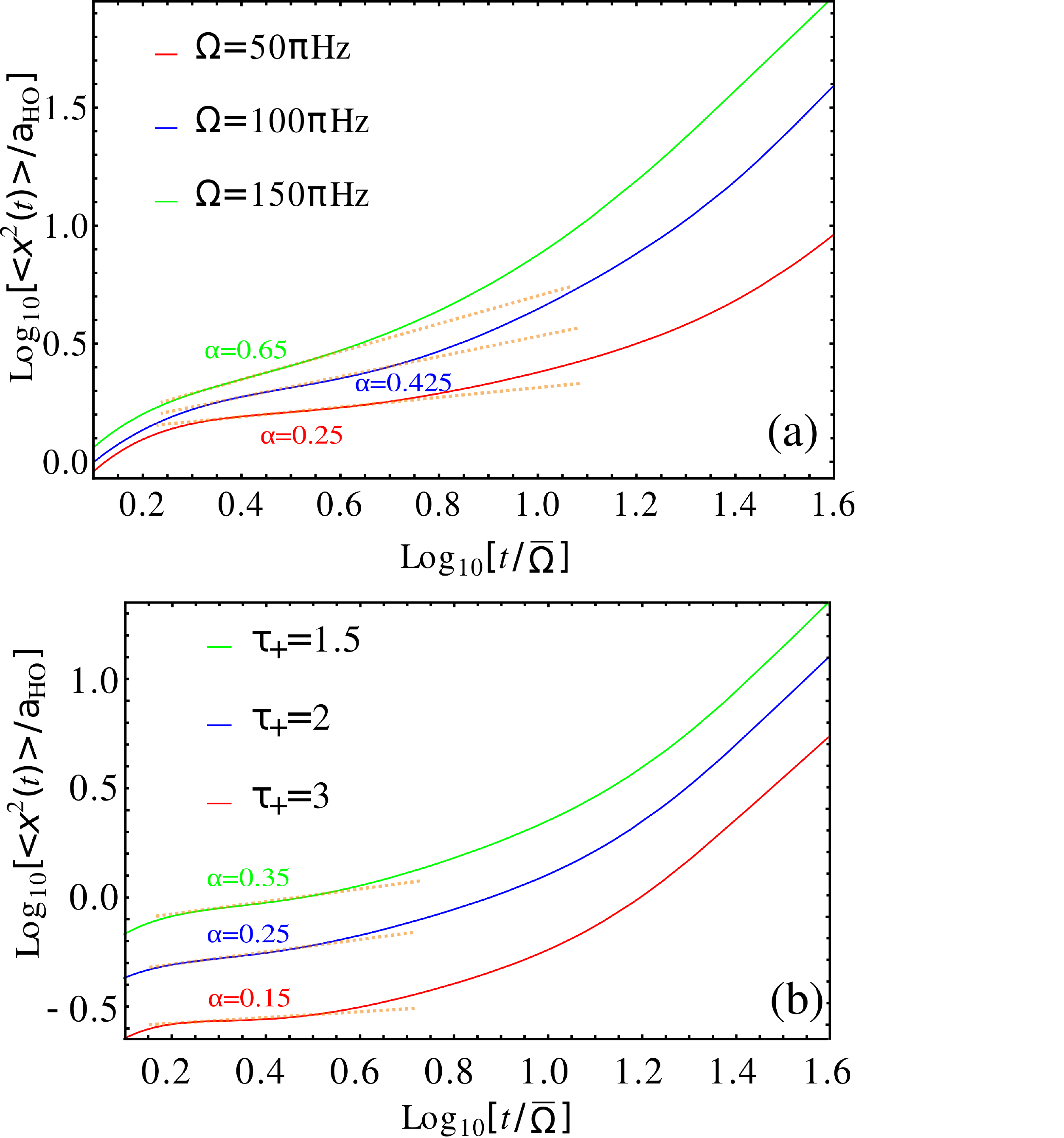}
%\tabularnewline
%\hline 
%\end{tabular}
\caption{Mean square displacement {\it vs} time for the case of coupling to the spin mode. A cutoff of $\Lambda=10\bar{\Omega}$ was used, where $\bar{\Omega}=1000\pi Hz$. In (a) we plot it for different coherent couplings $\Omega$ and in (b) for different couplings to the bath. The MSD shows three regimes, where it behaves approximately as $\mbox{MSD}\!\left(t\right)\propto t^{\alpha}$, and therefore linearly in log-log pots,  with a different slope given by the anomalous exponent $\alpha$: (i) an initial short time behavior, where $\alpha\approx 2$; (ii) a nontrivial transient subdiffusive behavior, where $\alpha<1$. We plot a dashed orange line as a guide to the eye, to illustrate the different slopes in this regime; (iii) a long time ballistic regime, with $\alpha= 2$. In (a) we show that, as $\Omega$  is reduced, the subdiffusive platteau enlarges and $\alpha$ gets smaller. In (b) we show that increasing  the couplings to the bath $\tau_+$, also enlarges the plateau and reduces $\alpha$. We consider Rb and K atoms for BEC and impurities, respectively. We use  $g=g_{12}=2.15\times10^{-37}J\cdot m$, density $n=7(\mu m)^{-1}$,  and impurity-BEC  $g_{\rm{IB}}=0.5\times10^{-37}J\cdot m$; We take $\tau_+=1$ in (a) and $ \Omega=100\pi Hz$ in (b). The BEC was assumed to be in the low temperature regime, i.e. when $\coth\left(\frac{\hbar \omega}{2k_{B}T}\right)\rightarrow1$ holds. \label{fig:fig2}}
\end{figure}

\section{Results: Mean square displacement}
\label{sec:results}

%\paragraph{\textbf{Mean square displacement}}

 With the Green propagator and the spectral density at hand, we are
now in a position to evaluate the MSD. This, as shown in \citep{2017Lampo}, is evaluated in the long time limit $\omega\ll\Lambda_{-},\Lambda$
as
\begin{align}\label{eq:MSD}
&\langle\left[x\left(t\right)-x\left(0\right)\right]^{2}\rangle=\mbox{MSD}\!\left(t\right)\!=\!G^2_{2}(t)\langle\dot{x}^{2}\left(0\right)\rangle\nonumber\\
&+\frac{1}{2}\!\int_{0}^{t}\!\!\!\!ds\!\!\int_{0}^{t}\!\!\!\!d\sigma G_{2}(s)G_{2}(\sigma)\!\left\langle\left\{\!B(s),B(\sigma)\!\right\}\right\rangle_{\rho_{B}},
\end{align}
where we assumed that the impurity-bath are initially in a product
state $\rho\left(0\right)=\rho_{B}\otimes\rho_{S}\left(0\right)$,
where $\rho_{B}$ is the thermal Gibbs state for the bath at temperature
$T$. The initial conditions of the impurity and bath oscillators are
then uncorrelated. Then, averages of the form $\left\langle \dot{x}\left(0\right)B(s)\right\rangle $
vanish. To treat the second term in Eq.~\eqref{eq:MSD}, we note that 
\begin{equation}
\left\langle \left\{ B(s),B(\sigma)\right\} \right\rangle _{\rho_{B}}=2\hbar\nu\left(s-\sigma\right),
\end{equation}
where $\nu\left(t\right)$ is defined as in Eq.~\eqref{eq:noise and damping kernel}.

In  case 1 (coupling to density mode) the spectral density reads
as in Eq.~\eqref{eq:Density mode spectral density}. Then,  the MSD behaves the same way as in~\citep{2017Lampo},
with the only difference of replacing $g\rightarrow g+g_{12}$. Hence
the impurity will again superdiffuse as 
\begin{equation}
\langle\left[x\left(t\right)-x\left(0\right)\right]^{2}\rangle=\left[\langle\dot{x}^{2}\left(0\right)\rangle+\frac{\tau_{-}\Lambda_{-}^{2}}{2}\right]\left(\frac{t}{\zeta}\right)^{2},\label{eq:MSD density mode}
\end{equation}
where $\zeta=1+\tau_{-}\Lambda_{-}$. Note that the superdiffusive
behavior $\langle x^{2}\left(t\right)\rangle\propto t^{2}$ appears
for both low temperature ($\coth\left(\hbar\omega/2k_{B}T\right)\approx1$)
and high temperature ($\coth\left(\hbar\omega/2k_{B}T\right)\approx2k_{B}T/\hbar\omega$)
limits. Hence from Eq.~\eqref{eq:MSD density mode} we see that, effectively,
the contribution of the Bogoliubov modes to the MSD behavior
in this case is just to modify the mass of the free particle. 

In case 2 (coupling to the spin mode), analytical  expressions
for the Eq. \eqref{eq:MSD} cannot be found. We remind again
that we are interested in the transient effects attributed to the
bath frequencies right above the band gap, after making the assumption
for $\omega=E_{gap}+\epsilon$ that $\omega\approx E_{\mathrm{gap}}$
i.e. $\epsilon\ll E_{\mathrm{gap}}$. In this case the Green function
reads as in Eq. \eqref{eq:approximate spectral density}, while the noise
kernel at low temperatures, where $\coth\left(\frac{\hbar \omega}{2k_{B}T}\right)\rightarrow1$,
can be shown to be equal to
\begin{align}
&\nu\left(t\right)=\tau_{+}\!\left(\Lambda\!\!-\!\!E_{\mathrm{gap}}\right)^{1.5}\\
&\times\!\left[\frac{2}{3}\cos\left(E_{\mathrm{gap}}t\right)\!F_{1,2}\!\left(\frac{3}{4};\frac{1}{2},\frac{7}{4};-\frac{1}{4}t^{2}\left(\!E_{\mathrm{gap}}\!\!-\!\!\Lambda\right)^{2}\!\right)\right.\nonumber\\
&+\frac{2}{5}t\left(E_{\mathrm{gap}}-\Lambda\right)\cos\left(E_{\mathrm{gap}}t\right)\nonumber\\
 &\left.\times F_{1,2}\left(\frac{5}{4};\frac{3}{2},\frac{9}{4};-\frac{1}{4}t^{2}\left(E_{\mathrm{gap}}-\Lambda\right)^{2}\right)\right],\nonumber
\end{align}
where 
\begin{equation}
F_{1,2}\left(\alpha;\beta,\gamma;z\right)=\sum_{n=0}^{\infty}\frac{\left(\alpha\right)_{n}}{\left(\beta\right)_{n}\left(\gamma\right)_{n}}\frac{z^{n}}{n!}.
\end{equation}

In this case, we were not able to obtain an
analytic solution for the MSD. We evaluate it numerically, with the results being valid for finite $\Omega$, as we used the simplified version of the spectral density, Eq.~\eqref{eq:approximate spectral density}. In all calculations we checked that all assumptions made are fulfilled. 

In Fig.~\ref{fig:fig2} we show the numerically evaluated  MSD as a function of time  according to Eq.~\eqref{eq:MSD} and different Rabi frequencies and interaction strengths. 
We remind that initially, $\langle\dot{x}^{2}\left(0\right)\rangle=0$.
In Fig.~\ref{fig:fig2} (a) we show how decreasing $\Omega$ both enlarges the duration of the subdiffusive plateau and reduces the anomalous exponent $\alpha$. We should note that the results are valid only for finite $\Omega$: since we use the simplifies spectral density, Eq.~\eqref{eq:approximate spectral density}, we are never able to describe the smooth transition to the cubic spectral density, which will show an smooth change to ballistic behavior for the whole range. Then, the effect  of reducing $\Omega$ is merely to reduce the gap, not to change the form of the spectral density. As a consequence, the plateau is enlarged. In Fig.~\ref{fig:fig2} (b) we show how the MSD varies as a function of the coupling
stength of the impurity to the BECs. Here, we observe that increasing the coupling strength results in more subdiffusive motion and an increase in the duration  of the subdiffusive plateau. 
Furthermore, note that, in Fig.~\ref{fig:fig2}, time  is measured
in units of the inverse of $\bar{\Omega}=1000\pi Hz$ and hence the transient subdiffusive phenomenon appears in time of the order of $ms$.    

In general, from the results presented in Fig.~\ref{fig:fig2}, we numerically find three regimes of behavior for the MSD, and in each regime it behaves as  $\mbox{MSD}\!\left(t\right)\propto t^{\alpha}$, where  $ \alpha $  is different at each regime. The exponent $ \alpha $ is known as the anomalous exponent.  In regime (i), there is an initial short time behavior where, as expected, the  MSD grows more or less ballistically with time. So here, $\alpha\approx 2$; In regime (ii), there is a plateau where $\alpha<1$. This is a transient subdiffusive behavior; Finally, in regime (iii), which is the long time behavior,  the impurity  superdiffuses  with $\alpha=2$. 
We interpret this behavior as follows: the impurity performs free motion initially. Then after interacting with the large frequencies of the bath, the impurity begins to perform a subdiffusive motion since it screens the part of the spectral density that depends on the square root of the bath modes frequencies. At long times, and after undergoing dissipation for some time, the impurity again effectively only interacts with the lower
frequencies of the bath which have zero effect on the motion of the
impurity and hence the impurity performs a ballistic motion.

Finally, we comment here that the results of the numerical integrations undertaken to obtain the MSD in this section, indeed appear to agree with the analytical calculations we performed in the previous section. This is to be understood in the following sense. The analytical results of the previous section, implied that the coupling of the impurity to the spin mode, results in the impurity interacting with a bath giving rise to a different spectral density, which intuitively one expects to give a different MSD behaviour. This is indeed what we observe in the results presented in Fig.~\ref{fig:fig2}.

\section{Conclusions}
\label{sec:conclusions}
In this work, we studied the diffusive behavior of an impurity immersed
in a coherently coupled two-component BEC, that interacts with both
of them through contact interactions. We showed how starting from
the standard Hamiltonian that would describe such a scenario, one
can recast the problem into
that of a quantum Brownian particle diffusing in a bath composed of
the Bogoliubov modes of the two-component BEC. We discussed the under certain assumptions and conditions required to obtain this description. 

We found that the main
difference of this scenario compared to that of the impurity being
coupled to a single BEC studied in~\citep{2017Lampo}, is that for
the scenario of the impurity being coupled differently to the two
BECs, namely coupled attractively to one of them and repulsively to
the other but with the same magnitude, results in the impurity being
coupled to the spin mode of the coherently coupled two-component BEC. This implies that its dynamics is determined by a qualitatively
different spectral density. In particular this new spectral density
is gapped and subohmic close to the gap. We demonstrate numerically,
that such a spectral density gives rise  to a transient subdiffusive
behavior. Furthermore, we show that this transient effect can be controlled
by the magnitude of the Rabi frequency, as well as by the strength
with which the impurity couples to the two BECs. These
can control the time duration for which this subdiffusive behavior
appears. A mechanism for inducing a transient controlled subdiffusion
in Brownian motion has been also proposed in~\citep{2016Spiechowicz},
but with a completely different way for achieving it and most importantly
not considering the system from a microscopic perspective. Moreover,
we comment that the setup we studied, thanks to the appearance of
this gapped subohmic spectral density, could also serve for simulating
quantum-optical phenomena, that could be seen for example in photonic
crystals, using instead cold atoms, as was proposed also in \citep{2011NavarreteBenlloch}
for the case of optical lattices. In addition, we note that our
studies could be extended to the scenario of having two impurities
in the coherently coupled two-component BEC, and study as in \citep{2019Charalambous},
the effects that the coupling to the spin mode could have on the bath-induced
entanglement between the two impurities. Finally, we could also study
the effect that this new gapped spectral density could have on the
functioning of the impurity as a probe to measure the temperature
of the two-component BEC, as in \citep{2019Mehboudi}. Last but not least, it should be noted here that, if one considers the scenario of attractive two-body coupling, i.e. $g_{12}<0$, and includes the Lee-Huang-Yang corrections to the Hamiltonian, one will obtain the scenario of quantum droplets studied theoretically in \cite{2015Petrov} and recently proven experimentally in \cite{2018Cabrera} . This we expect to lead in different interesting dynamics for the immersed impurity in the two-component coherently coupled BEC.       

\acknowledgments 
We (M.L. group) acknowledge the Spanish Ministry MINECO (National Plan
15 Grant: FISICATEAMO No. FIS2016-79508-P, FPI), the Ministry of Education of Spain (FPI Grant BES-2015-071803), EU FEDER, European Social Fund, Fundació Cellex, Generalitat de Catalunya (AGAUR Grant No. 2017 SGR 1341 and CERCA/Program), ERC AdG OSYRIS and NOQIA, EU FETPRO QUIC, and the National Science Centre, Poland-Symfonia Grant No. 2016/20/W/ST4/00314. MAGM acknowledges funding from the Spanish Ministry of Education and Vocational Training (MEFP) through the Beatriz Galindo program 2018 (BEAGAL18/00203).
 
%\bibliographystyle{iopart-num}

%\bibliographystyle{plainnat}
%\bibliography{Biblio2BEC.bib}

\bibliographystyle{unsrtnat}
\bibliography{Biblio2BEC.bib}

\end{document}